\documentclass[oneside,english]{amsart}
\usepackage[T1]{fontenc}
\usepackage[latin9]{inputenc}
\usepackage{babel}
\usepackage{mathtools}
\usepackage{amstext}
\usepackage{amsthm}
\usepackage{amssymb}
\usepackage[pdfusetitle,
 bookmarks=true,bookmarksnumbered=false,bookmarksopen=false,
 breaklinks=false,pdfborder={0 0 1},backref=false,colorlinks=false]
 {hyperref}
\usepackage{stmaryrd}

\makeatletter
\numberwithin{equation}{section}
\numberwithin{figure}{section}
\theoremstyle{plain}
\newtheorem{thm}{\protect\theoremname}
\theoremstyle{definition}
\newtheorem{defn}[thm]{\protect\definitionname}
\theoremstyle{plain}
\newtheorem{prop}[thm]{\protect\propositionname}
\theoremstyle{remark}
\newtheorem*{rem*}{\protect\remarkname}
\theoremstyle{plain}
\newtheorem{lem}[thm]{\protect\lemmaname}
\theoremstyle{definition}
\newtheorem*{defn*}{\protect\definitionname}
\theoremstyle{plain}

\theoremstyle{remark}
\newtheorem{rem}[thm]{\protect\remarkname}

\DeclareMathOperator{\cost}{cost}
\DeclareMathOperator{\depth}{depth}
\DeclareMathOperator{\leaves}{leaves}
\newcommand{\sem}[1]{\llbracket #1 \rrbracket}
\newcommand{\lean}[1]{{\ttfamily\small #1}}
\newcommand{\bigstep}{\Downarrow}
\makeatother

\usepackage{listings}
\usepackage{booktabs}
\usepackage{array}
\providecommand{\definitionname}{Definition}
\providecommand{\corollaryname}{Corollary}
\providecommand{\lemmaname}{Lemma}
\providecommand{\propositionname}{Proposition}
\providecommand{\remarkname}{Remark}
\providecommand{\theoremname}{Theorem}

\begin{document}
\title[CrossCheck I: Cost Bounds as Constrained Optimization]{CrossCheck I: Cost Bounds of Imperative Programs as Constrained Optimization, with a Sound Cost Logic and a Verified Certificate Checker}
\author{Luca Dall'Olio}
\thanks{Alter Ego Engineering, 5 rue du Docteur Lequeux, 92330 Sceaux, France. Email: \texttt{ld@aego.ai}. ORCID: \href{https://orcid.org/0009-0006-8839-5909}{0009-0006-8839-5909}. This preprint: \href{https://doi.org/10.5281/zenodo.24444576}{doi:10.5281/zenodo.24444576}.}
\begin{abstract}
Static analysis of imperative programs often relies on iterative fixed-point approximations or concrete path exploration. We show that structured imperative control flow---while loops, linear recursion and branching recursion---can be mapped into constrained optimization problems over the free monoid of execution traces and the free algebra of computation trees. Ordering traces and trees by an intensional \textit{order-by-composition} ($\sqsubseteq$), halting is characterized as finding the minimal trace or tree that satisfies the exit condition, and we prove that this characterization coincides with the standard big-step semantics. Exact state recurrences are undecidable in general, so we relax them into cost envelopes, using ranking potentials and the tropical $(\max,+)$ reading of the timing schema, and obtain sound upper bounds on execution cost. The bounds compose over whole programs in a cost logic for a small language with nondeterministic actions, procedures, recursion and assumed contracts, which we prove sound for terminating and non-terminating executions. A linear instance of the logic has certificates---programs annotated with invariants, potentials and Farkas multipliers---and a checker that validates them with integer arithmetic only; an accepted certificate bounds the cost of every execution. All results, the checker and the worked examples are machine-checked in Lean~4. A companion paper describes a Clang plugin that produces such certificates and its evaluation.
\end{abstract}

\maketitle

\section{Introduction}

Static analysis of imperative programs has traditionally followed two divergent trajectories:
concrete path exploration (such as bounded model checking and symbolic execution) and semantic
over-approximation (such as abstract interpretation). Tools operating at the source or bytecode level
rely on traversing control-flow graphs (CFGs) or computing iterative fixed points over abstract
domains \cite{cousot1977abstract}. While effective for detecting specific property violations,
these approaches do not directly provide closed-form analytical descriptions of execution resources,
and may suffer from state-space explosion or precision loss during domain widening.

In safety-critical and hard real-time systems, the central verification question is rarely whether
one can predict the exact contents of every memory cell upon termination; rather, it is whether the
program is guaranteed to terminate within a strict execution budget
(Worst-Case Execution Time, or WCET) \cite{wilhelm2008worst}. Classical approaches to this problem,
such as the Implicit Path Enumeration Technique (IPET) \cite{li1995ilp}, analyze execution frequencies
over control-flow graphs and require loop bounds as an input, obtained from annotations or from a
separate bound analysis \cite{sinn2014complexity,alias2010multi}; recursion, and branching recursion
in particular, is usually excluded or handled by dedicated annotations.

In this paper, we introduce \textbf{CrossCheck}, a framework that transforms structured imperative control
flow---iterative loops, linear recursion, and non-linear branching recursion---into
constrained optimization problems without step-by-step simulation or unrolling.
The framework rests on a dual-layer semantic architecture:
\begin{enumerate}
    \item \textbf{An Intensional Layer:} Execution is modeled as words in the free instruction
    monoid $(\Sigma^*, \cdot, \epsilon)$ and as computation trees in a free term algebra $\mathcal{T}_\Sigma$.
    Both structures carry an \textit{order-by-composition} relation ($\sqsubseteq$), a well-founded partial
    order along which execution cost (instruction or work count) is additive and monotone.
    \item \textbf{An Extensional Layer:} A semantic valuation maps traces to state transformers
    over a pointed execution domain $S_\bot$.
\end{enumerate}

Under this framework, loop execution and recursive unwinding are characterized as finding the
\textit{minimal exiting trace or computation tree} under $\sqsubseteq$ that satisfies the
exit condition. For loops whose updates are translations, this optimization problem yields exact
closed-form terminal states and cycle counts simultaneously.

For general programs, exact state calculation is undecidable: this already follows from the
undecidability of the halting problem, and by Matiyasevich's Theorem \cite{matiyasevich1970enumerable}
it persists for loops restricted to polynomial integer arithmetic. The extensional state equations
are therefore relaxed into cost envelopes. By combining the timing schema of structured programs
\cite{shaw1989reasoning,puschner1989calculating}, read in the tropical $(\max, +)$ semiring \cite{baccelli1992synchronization},
with Lyapunov-style ranking potentials
\cite{floyd1967assigning,podelski2004transition}, CrossCheck derives sound upper bounds on execution cost.

\subsection*{Positioning}
The individual ingredients of this work are classical, and we do not claim otherwise. The minimal iteration count
$\bar{n} = \min\{n \mid \neg P(f^n(s_0))\}$ is an instance of Kleene's minimization ($\mu$) operator; ranking functions
date back to Floyd \cite{floyd1967assigning}, and their use to derive complexity bounds is well established
\cite{alias2010multi,sinn2014complexity}; the rule that bounds a conditional by its most expensive
branch and a loop by its iteration bound times its body is the timing schema of Shaw and of Puschner and Koza
\cite{shaw1989reasoning,puschner1989calculating}, which we merely phrase in max-plus algebra \cite{baccelli1992synchronization}. The contribution of CrossCheck is the \emph{uniform packaging} of these
ingredients: a single order-theoretic formulation (minimality under $\sqsubseteq$) that covers loops, linear recursion
and branching recursion alike, that carries cost and state at the same time, and whose soundness with respect to a
standard big-step semantics is machine-checked end to end.

\subsection*{Contributions}
The primary contributions of this work are:
\begin{itemize}
    \item \textbf{Order-by-Composition:} We use the prefix order $\sqsubseteq$ on the free trace monoid
    and the corresponding order on computation trees, in which a leaf may be replaced by any tree
    (Courcelle's order on trees \cite{courcelle1983trees}), prove the Minimal Exiting Trace Theorem via
    the well-ordering of $\mathbb{N}$, and prove that the minimizer coincides with the result of the standard
    big-step semantics of while loops (Theorems~\ref{thm:minimal-trace} and~\ref{thm:extensional-equivalence}).
    \item \textbf{Generalization to Recursion:} We extend the characterization to linear recursion
    (Proposition~\ref{prop:linrec}) and to branching recursion, where the execution tree is the \emph{least} tree, with respect to the
    tree order-by-composition, whose leaves all exit (Proposition~\ref{prop:tree-optimization}); uniform recursive
    unfolding corresponds to tree exponentiation ($B^{\otimes d}$).
    \item \textbf{Cost Envelopes:} We relax the exact problems into sound upper bounds: tropical branch collapsing,
    a Lyapunov iteration bound with termination (Theorem~\ref{thm:lyapunov-bound}), and a depth envelope for branching
    recursion (Proposition~\ref{prop:tree-envelope}).
    \item \textbf{A Sound Cost Logic:} For a small imperative language with nondeterministic atomic actions we give a
    proof system for cost bounds whose loop rule takes these certificates as premises, and prove it sound for
    terminating and non-terminating executions (Theorem~\ref{thm:cost-logic-soundness}); the system has assumed
    triples for code that is not analyzed and a rule for recursive procedures, and covers bounds that depend on the
    initial state, with a call rule for function summaries (Proposition~\ref{prop:summary-call}).
    \item \textbf{A Verified Certificate Checker:} For the linear instance of the logic we define certificates,
    programs annotated with invariants, potentials and Farkas multipliers, and a checker that uses integer
    arithmetic only; a certificate that it accepts bounds the cost of every execution of its program
    (Theorem~\ref{thm:checker-soundness}). Bounds may be affine or quadratic in the parameters of a function, and calls are
    checked against the callee's summary and against assumed contracts instead of being inlined
    (Theorem~\ref{thm:program-soundness}). A certificate can also be validated against an independent
    description of the source in a small fragment of C, so that its bound holds for that description
    without trusting the producer of the certificate (Theorem~\ref{thm:validated}).
    \item \textbf{Interrupts and Fork-Join:} From the certified bounds of a computation and of the interrupt
    handlers that preempt it, a response-time bound (Theorem~\ref{thm:response-time}), which the checker
    verifies; and Graham's bound for a fork--join block of independent tasks (Proposition~\ref{prop:graham}).
    Both results are classical; they are mechanized here and connected to the certificates.
    \item \textbf{Machine-Checked Soundness:} All of the above, together with the worked examples, are formalized
    in Lean~4 \cite{demoura2021lean4} on top of mathlib \cite{mathlib2020}, without \lean{sorry} and without any axiom
    beyond the three standard axioms of Lean's logic (Section~\ref{sec:mechanization}).
    \item \textbf{Constraint Generation:} We describe operational rules that emit recurrence constraint systems from
    imperative code, and the design of an SMT-based synthesis of linear ranking potentials for compile-time
    enforcement of timing budgets (Section~\ref{sec:application}).
\end{itemize}

A companion paper \cite{crosscheckTool} describes a Clang plugin that implements the relaxed analysis for C and C++ functions, emits certificates for the checker of Section~\ref{subsec:checker}, and compares its results with five existing analyzers on established benchmarks; a third paper \cite{crosscheckBinary} certifies cycle bounds of compiled machine code, its loop bounds included. This paper contains no measurement.

\section{Related Work}
\label{sec:related}

\paragraph{Cost and bound analysis.}
Automatic derivation of symbolic resource bounds goes back to Wegbreit \cite{wegbreit1975mechanical}.
Modern analyzers derive bounds from ranking functions, counters and invariants
\cite{gulwani2009speed,alias2010multi,sinn2014complexity,brockschmidt2016analyzing,lommen2022automatic}, from cost
relations solved into closed forms \cite{albert2011closed}, or from potential-based
amortized analysis \cite{hoffmann2012multivariate,carbonneaux2015compositional}; see \cite{hoffmann2022two} for a survey of
the latter. Our Lyapunov bound (Theorem~\ref{thm:lyapunov-bound})
is the single-loop core shared by these methods; we do not improve on their precision (in particular we perform no
amortization across nested loops), but we embed the bound in a semantics where cost, trace and state are
defined by the same minimization problem.

\paragraph{Ranking function synthesis.}
Linear ranking functions can be synthesized by linear programming through Farkas' Lemma
\cite{colon2001synthesis,podelski2004complete}; transition invariants \cite{podelski2004transition} generalize the
approach to disjunctive well-founded relations. Ben-Amram and Genaim \cite{benamram2014ranking} settled the complexity of
the problem: deciding the existence of a linear ranking function is polynomial over the rationals and coNP-complete over
the integers, which is why the Farkas-based method is sound but incomplete for integer loops.
Section~\ref{subsubsec:smt-synthesis} adopts the linear-template method unchanged.

\paragraph{Loop acceleration.}
Replacing a loop by a closed form of its $n$-fold iterate is known as acceleration \cite{bardin2008fast}; recent work
combines acceleration techniques modularly \cite{frohn2020calculus}, and closed forms of linear and polynomial loops, with decision procedures for their safety and termination, are computed in \cite{kincaid2019closed}. Our exact
(non-relaxed) treatment of loops is an acceleration together with a minimization over $n$.

\paragraph{WCET analysis.}
Source-level timing analysis started with the timing schema of Shaw \cite{shaw1989reasoning} and of Puschner and Koza
\cite{puschner1989calculating}, which assign to each structured statement a time bound computed from the bounds of its
parts: a sum for sequences, a maximum for alternatives, and the iteration bound times the body for loops. The cost
rules of Section~\ref{subsec:asymptotic-relaxation} are this schema; what we add is its derivation from, and its
machine-checked soundness with respect to, a trace semantics. Puschner and Koza also let the programmer bound how many times a marked point is passed within a loop (markers and scopes) and the total number of iterations of a sequence of loops, constraints on execution counts of the kind that implicit path enumeration expresses as linear constraints \cite{li1995ilp}.
Industrial WCET analyzers combine loop-bound analysis, micro-architectural timing analysis and IPET path analysis
\cite{wilhelm2008worst,li1995ilp}. CrossCheck addresses only the first and third ingredients, at the level of an abstract
cost norm; see Section~\ref{sec:limitations}.

\paragraph{Mechanized cost analysis.}
A machine-checked loop-bound estimation for WCET analysis was developed
in Coq by Blazy, Maroneze and Pichardie \cite{blazy2013formal} and extended to a formally verified WCET estimation tool
integrated in the CompCert compiler \cite{maroneze2014formally}; certified resource bounds for C programs were obtained
in \cite{carbonneaux2015compositional}. Program logics for running time have been formalized in Isabelle/HOL
\cite{haslbeck2018hoare} and in Coq, with asymptotic bounds \cite{gueneau2018fistful}. Atkey \cite{atkey2010amortised} embeds amortised resource analysis in separation logic for a small language modelled on Java bytecode, proves the logic sound in Coq, extracts a certified generator of verification conditions, and infers the resource annotations by linear programming, a division of labour close to ours between an untrusted search and a checked proof. Compared with these, our
mechanization is smaller in scope: it covers the semantic characterization and the bounds of this paper, not a
verified analyzer for a real programming language.

\paragraph{Semantics.}
Big-step (natural) semantics is due to Kahn \cite{kahn1987natural}; fixed-point semantics of loops and recursion is
standard \cite{manna1974mathematical,winskel1993formal}. Big-step semantics relates only terminating executions; divergence
can be captured coinductively \cite{leroy2009coinductive}, whereas here it appears as the infeasibility of the
minimization problem. Our traces are plain words of a free monoid; they are unrelated to Mazurkiewicz traces
\cite{diekert1995traces}, which are elements of partially commutative monoids.

\section{Foundations}
\label{sec:foundations}

We recall basic definitions and fix the notation. For background on operational and
denotational semantics see \cite{winskel1993formal}.

\subsection{States, traces and valuation}

\begin{defn}[State Space and Partial State Transitions]
\label{def:state-transition}
Let $S$ be a state space representing memory configurations (e.g., $S \coloneqq \text{Var} \to \text{Val}$). A \textit{partial state transition} on $S$ is a partial function $f : S \rightharpoonup S$, i.e., a relation $f \subseteq S \times S$ such that
\[
\forall\, s, s', s'' \in S, \quad (s, s') \in f \land (s, s'') \in f \implies s' = s''.
\]
We denote by $\bot$ an undefined execution outcome (an unhandled dynamic fault) and extend $S$ to the pointed execution domain $S_\bot \coloneqq S \cup \{\bot\}$. The transition $f$ extends to a total, strict operator $f_\bot : S_\bot \to S_\bot$:
\[
f_\bot(s) \coloneqq \begin{cases}
f(s) & \text{if } s \in S \text{ and } f \text{ is defined at } s, \\
\bot & \text{otherwise}.
\end{cases}
\]
\end{defn}

\begin{defn}[Traces, Iteration and Valuation]
\label{def:trace-functional}
Let $\Sigma$ be an alphabet of primitive atomic instructions, and let $\Sigma^*$ denote the free monoid of finite execution traces under concatenation $\cdot$, with identity element $\epsilon$ (the empty trace).

\smallskip\noindent
1. \textbf{Intensional iteration.} The iteration functional $\Phi : \Sigma^* \times \mathbb{N} \to \Sigma^*$ is defined by
\[
\Phi(t, n) \coloneqq t^n = \begin{cases}
\epsilon & \text{if } n = 0, \\
t^{n-1} \cdot t & \text{if } n \ge 1.
\end{cases}
\]

\smallskip\noindent
2. \textbf{Extensional valuation.} Let every instruction $a \in \Sigma$ be interpreted by a partial state transition $I(a) : S \rightharpoonup S$. The valuation $\sem{\cdot} : \Sigma^* \to (S_\bot \to S_\bot)$ is the unique map such that
\[
\sem{\epsilon} = \text{Id}_{S_\bot}, \qquad \sem{a} = I(a)_\bot \ (a \in \Sigma), \qquad \sem{t_1 \cdot t_2} = \sem{t_2} \circ \sem{t_1}.
\]
\end{defn}

By freeness of $\Sigma^*$ the valuation is well defined; it is a monoid homomorphism into the strict operators on $S_\bot$ under reverse composition. In particular
\begin{equation}
\label{eq:iter-hom}
\sem{t^n} = \sem{t}^{\,n} \qquad \text{for all } t \in \Sigma^*,\ n \in \mathbb{N},
\end{equation}
by induction on $n$ (\lean{evalTrace\_\allowbreak concat}, \lean{evalTrace\_\allowbreak iter}).

\subsection{Cost norms}

\begin{defn}[Cost Norm]\label{def:subadditive-norm}
A function $\|\cdot\| : \Sigma^* \to \mathbb{N}$ is a \textit{subadditive cost norm} if $\|\epsilon\| = 0$ and
$\|u \cdot v\| \le \|u\| + \|v\|$ for all $u, v \in \Sigma^*$. Given instruction weights $\omega : \Sigma \to \mathbb{N}$, the
\textit{weighted cost}
\[
\|a_1 a_2 \cdots a_k\|_\omega \coloneqq \sum_{j=1}^{k} \omega(a_j)
\]
is an additive cost norm: $\|u \cdot v\|_\omega = \|u\|_\omega + \|v\|_\omega$, hence $\|t^n\|_\omega = n \cdot \|t\|_\omega$. The \textit{instruction count} $\cost(t) \coloneqq |t|$ is the weighted cost with unit weights.
\end{defn}

The weight $\omega(a)$ is meant to be an upper bound on the execution time of instruction $a$ on the target platform; unit weights give a platform-independent step count. All exact identities below are stated for $\cost$ and hold verbatim for every weighted cost; all upper bounds hold for every subadditive cost norm, since subadditivity gives $\|t_1 \cdots t_n\| \le \sum_j \|t_j\|$.

\subsection{Order-by-composition}

\begin{defn}[Intensional Order-by-Composition]\label{def:order-by-composition}
The \textit{order-by-composition} (prefix order) $\sqsubseteq$ on the trace monoid $\Sigma^*$ is defined by
\[
u \sqsubseteq w \iff \exists\, v \in \Sigma^* \quad \text{such that} \quad w = u \cdot v.
\]
\end{defn}
This is the usual prefix order on words; the name records the reading used below, in which $w$ continues the execution $u$.

\begin{prop}[Properties of Intensional Order-by-Composition]
\label{prop:poset-properties}
The relation $\sqsubseteq$ is a partial order on $\Sigma^*$ with least element $\epsilon$, and it is well-founded. Moreover $\cost$ is additive, $\cost(u \cdot v) = \cost(u) + \cost(v)$ and $\cost(t^n) = n \cdot \cost(t)$, and monotone: $u \sqsubseteq w \implies \cost(u) \le \cost(w)$.
\end{prop}

\begin{proof}
Reflexivity holds since $w = w \cdot \epsilon$, and transitivity follows from associativity of concatenation. For antisymmetry, $u \sqsubseteq w$ and $w \sqsubseteq u$ give $|u| \le |w| \le |u|$, so the extension $v$ in $w = u \cdot v$ has length $0$ and $u = w$. Additivity of length gives the cost identities and monotonicity; well-foundedness follows because a strict prefix is strictly shorter.
\end{proof}

\begin{prop}[Monotonicity of Trace Iteration]
\label{prop:n-monotony}
For any trace $t$ and $n \le m$ we have $t^n \sqsubseteq t^m$. If $t \neq \epsilon$, the converse also holds:
\[
t^n \sqsubseteq t^m \iff n \le m.
\]
Hence, for $t \ne \epsilon$, $n \mapsto t^n$ is an order isomorphism from $(\mathbb{N}, \le)$ onto the chain $(\{t^n\}_{n \in \mathbb{N}}, \sqsubseteq)$.
\end{prop}

\begin{proof}
If $n \le m$ then $t^m = t^n \cdot t^{m-n}$. Conversely, $t^n \sqsubseteq t^m$ implies $n \cdot |t| \le m \cdot |t|$ by Proposition~\ref{prop:poset-properties}, and $|t| > 0$ gives $n \le m$.
\end{proof}

Throughout, $P : S \to \mathbb{B}$ denotes the \emph{continuation} predicate of a loop (the loop keeps running while $P$ holds); we say that $u \in S_\bot$ \emph{exits} if $u \in S$ and $P(u) = \text{false}$. A faulting state $\bot$ never exits.

\begin{thm}[Minimal Exiting Trace Existence]
\label{thm:minimal-trace}
Let $t \in \Sigma^*$ be the trace of a loop body, $P$ the continuation predicate and $s_0 \in S$ an initial state. If the set
\[
N \coloneqq \{ n \in \mathbb{N} \mid \sem{t^n}(s_0) \text{ exits} \}
\]
is non-empty, it has a unique minimum $\bar{n}$. The trace $\bar{t} \coloneqq t^{\bar{n}}$ is the least element, with respect to $\sqsubseteq$, of $\{ t^n \mid n \in N \}$, and $\cost(\bar{t}) = \bar{n} \cdot \cost(t) \le \cost(t^n)$ for all $n \in N$.
\end{thm}

\begin{proof}
$N$ is a non-empty subset of $\mathbb{N}$, so it has a unique minimum $\bar{n}$ by the well-ordering principle. For $n \in N$ we have $\bar n \le n$, hence $t^{\bar{n}} \sqsubseteq t^{n}$ by Proposition~\ref{prop:n-monotony}; the cost statement is Proposition~\ref{prop:poset-properties}.
\end{proof}

\begin{rem}
The number $\bar{n}$ is the value of Kleene's minimization operator, $\bar{n} = \mu n.\, [\sem{t^n}(s_0) \text{ exits}]$. Theorem~\ref{thm:minimal-trace} adds that the minimization can equivalently be read in the trace poset, where it simultaneously minimizes every monotone cost.
\end{rem}

\subsection{Bodies with data-dependent paths}

A loop body containing conditionals does not emit one fixed trace: the instructions executed in an iteration depend on the current state.

\begin{defn}[Path Selector and Executed Trace]\label{def:path-selector}
A \textit{path selector} is a map $\tau : S \to \Sigma^*$ assigning to each state the trace emitted by one execution of the body from that state. Assume $\sem{\tau(s)}(s) \in S$ for the states of interest and write $f(s) \coloneqq \sem{\tau(s)}(s)$ for the induced state update. The trace executed by $n$ iterations from $s$ is
\[
\rho_n(s) \coloneqq \tau(s_0) \cdot \tau(s_1) \cdots \tau(s_{n-1}), \qquad s_k \coloneqq f^k(s).
\]
A body is \textit{straight-line} if $\tau$ is constant, $\tau(s) = t$; then $\rho_n(s) = t^n$.
\end{defn}

For a conditional $\text{if } c \text{ then } t_{\text{then}} \text{ else } t_{\text{else}}$ with guard trace $t_c$, the path selector is
\begin{equation}
\label{eq:cond-selector}
\tau_{\text{if}}(s) \coloneqq t_c \cdot \begin{cases} t_{\text{then}} & \text{if } c(\sem{t_c}(s)) \text{ holds}, \\ t_{\text{else}} & \text{otherwise.} \end{cases}
\end{equation}
Theorem~\ref{thm:minimal-trace} carries over verbatim with $f^n(s_0)$ in place of $\sem{t^n}(s_0)$ and $\rho_n(s_0)$ in place of $t^n$, since $\rho_n(s_0) \sqsubseteq \rho_m(s_0)$ whenever $n \le m$.

\begin{prop}[Worst-Case Step Bound]\label{prop:worst-case-step}
Let $\|\cdot\|$ be a subadditive cost norm and $W \in \mathbb{N}$ such that $\|\tau(s)\| \le W$ for all $s \in S$. Then $\|\rho_n(s)\| \le n \cdot W$ for all $n$ and $s$.
\end{prop}

\begin{proof}
By induction on $n$, using $\rho_{n+1}(s) = \tau(s) \cdot \rho_n(f(s))$ and subadditivity.
\end{proof}

\subsection{Nested and sequential loops}

\begin{prop}[Cost of Nested and Sequential Composition]\label{prop:nested-cost}
Let $t_1, t_2 \in \Sigma^*$ and $n, m, n_1, n_2 \in \mathbb{N}$.
\begin{enumerate}
    \item \textbf{Nested loops.} If each of $n$ outer iterations executes the straight-line part $t_1$ followed by $m$ iterations of an inner body $t_2$, the executed trace is $(t_1 \cdot t_2^m)^n$ and
    \[
    \cost\left((t_1 \cdot t_2^m)^n\right) = n \cdot \cost(t_1) + n \cdot m \cdot \cost(t_2).
    \]
    \item \textbf{Sequential loops.} The trace $t_1^{n_1} \cdot t_2^{n_2}$ of two loops in sequence has cost $n_1 \cdot \cost(t_1) + n_2 \cdot \cost(t_2)$, and the first loop is a prefix of the whole: $t_1^{n_1} \sqsubseteq t_1^{n_1} \cdot t_2^{n_2}$.
\end{enumerate}
\end{prop}

\begin{proof}
Both statements follow from additivity of $\cost$ (Proposition~\ref{prop:poset-properties}) and Definition~\ref{def:order-by-composition}.
\end{proof}

In general the inner iteration count depends on the state at each outer iteration; this case is covered by the path-selector formulation (Definition~\ref{def:path-selector}) and bounded in Section~\ref{subsubsec:compositional-wcet}.

\subsection{The tropical algebra of control flow}

\begin{prop}[Branch Rule]\label{prop:tropical-dominance}
Let $\|\cdot\|$ be a subadditive cost norm. For the conditional path selector \eqref{eq:cond-selector} and every state $s \in S$,
\[
\|\tau_{\text{if}}(s)\| \le \|t_c\| + \max\left(\|t_{\text{then}}\|, \|t_{\text{else}}\|\right).
\]
\end{prop}
\begin{proof}
Execution takes exactly one branch, so by subadditivity $\|\tau_{\text{if}}(s)\|$ is at most $\|t_c\| + \|t_{\text{then}}\|$ or at most $\|t_c\| + \|t_{\text{else}}\|$, and both are below the stated bound.
\end{proof}

This is the rule for alternatives of the timing schema \cite{shaw1989reasoning,puschner1989calculating}. Together with subadditivity for sequencing, it says that costs of structured code compose in the tropical semiring: sequencing is $+$ and branching is $\max$ (Section~\ref{subsubsec:tropical-semiring}).

\subsection{Big-step semantics and extensional equivalence}

Which power $t^n$ of the body trace a loop actually executes is determined by the extensional layer. To state this precisely we fix the standard big-step semantics \cite{kahn1987natural,winskel1993formal} of the loop with body $t$ and continuation predicate $P$, as the least relation $\bigstep$ on $S \times S$ closed under the rules
\begin{equation}
\label{eq:bigstep-while}
\frac{P(s) = \text{false}}{s \bigstep s}
\qquad\qquad
\frac{P(s) = \text{true} \quad \sem{t}(s) = s' \in S \quad s' \bigstep s''}{s \bigstep s''}
\end{equation}

\begin{thm}[Extensional Equivalence and Solution Soundness]
\label{thm:extensional-equivalence}
Let $t \in \Sigma^*$, $P : S \to \mathbb{B}$ and $s_0, \bar{s} \in S$. Then $s_0 \bigstep \bar{s}$ if and only if there exists $\bar{n} \in \mathbb{N}$ such that
\begin{enumerate}
    \item $\sem{t^{\bar{n}}}(s_0) = \bar{s}$ and $P(\bar{s}) = \text{false}$, and
    \item for every $k < \bar{n}$, $\sem{t^{k}}(s_0) \in S$ and $P(\sem{t^{k}}(s_0)) = \text{true}$.
\end{enumerate}
Such an $\bar{n}$ is unique; it is the minimum of Theorem~\ref{thm:minimal-trace}. Consequently the terminal state is the solution of the extensional recurrence system
\[
s_{k+1} = \sem{t}(s_k) \quad (0 \le k < \bar{n}), \qquad \bar{s} = s_{\bar{n}}, \qquad \bar{n} = \min \{ n \mid \neg P(s_n) \},
\]
and the big-step semantics is deterministic.
\end{thm}

\begin{proof}
($\Rightarrow$) By induction on the derivation of $s_0 \bigstep \bar{s}$. For the first rule take $\bar{n} = 0$. For the second rule, the induction hypothesis gives a count $n'$ for $s'$; take $\bar{n} = n' + 1$ and use $\sem{t^{k+1}}(s_0) = \sem{t^{k}}(\sem{t}(s_0)) = \sem{t^k}(s')$, which follows from \eqref{eq:iter-hom}. ($\Leftarrow$) By induction on $\bar{n}$. If $\bar{n} = 0$ then $\bar{s} = s_0$ and the first rule applies. Otherwise (2) with $k = 0$ gives $P(s_0) = \text{true}$; moreover $\sem{t}(s_0) \neq \bot$, since strictness would force $\sem{t^{\bar{n}}}(s_0) = \bot$. The induction hypothesis applies to $s' = \sem{t}(s_0)$ with count $\bar{n} - 1$, and the second rule concludes. Uniqueness: if $n < n'$ both satisfied (1)--(2), then $\sem{t^n}(s_0)$ would both falsify $P$ by (1) and satisfy it by (2).
\end{proof}

If no such $\bar{n}$ exists, the loop has no big-step derivation: it either faults (some $\sem{t^k}(s_0) = \bot$ is reached while $P$ held) or diverges.

\begin{rem*}[Fixed-point reading]
The chain $\epsilon \sqsubseteq t \sqsubseteq t^2 \sqsubseteq \cdots$ is the Kleene chain of the unrolling operator $x \mapsto t \cdot x$ on finite and infinite words ordered by prefix; its least upper bound, the least fixed point, is the infinite word $t^\omega$ when $t \ne \epsilon$ \cite{winskel1993formal}. An executed trace is a finite prefix $t^{\bar{n}}$ of it, and a diverging loop emits all of it. We do not use this reading in the sequel.
\end{rem*}

\subsection{Computation trees}

\begin{defn}[Computation Trees and Grafting]\label{def:tree-algebra}
Fix a branching factor $b \ge 2$. The set $\mathcal{T}_\Sigma$ of \textit{computation trees} is defined inductively:
\begin{enumerate}
    \item a leaf $\epsilon$ is a computation tree (a base-case call);
    \item if $T_1, \dots, T_b$ are computation trees, so is the node $g(T_1, \dots, T_b)$ (a call that branches into $b$ recursive calls).
\end{enumerate}
The \textit{grafting} $T_1 \otimes T_2$ substitutes $T_2$ for every leaf of $T_1$. More generally, a \textit{leaf substitution} replaces each leaf of $T_1$ by an arbitrary computation tree, possibly different for each leaf.
\end{defn}

Grafting is associative with unit $\epsilon$. We write $\cost(T)$ for the number of nodes of $T$ (leaves included), $\leaves(T)$ for its number of leaves, and $\depth(T)$ for its height, with $\depth(\epsilon) = 0$.

\begin{defn}[Tree Order-by-Composition]\label{def:tree-order}
The \textit{tree order-by-composition} $\sqsubseteq_{\text{tree}}$ on $\mathcal{T}_\Sigma$ is defined by
\[
T_1 \sqsubseteq_{\text{tree}} T_2 \iff T_2 \text{ is obtained from } T_1 \text{ by a leaf substitution.}
\]
Equivalently, it is the least relation such that $\epsilon \sqsubseteq_{\text{tree}} T$ for every $T$, and
\[
g(T_1, \dots, T_b) \sqsubseteq_{\text{tree}} g(T_1', \dots, T_b') \quad \text{whenever } T_i \sqsubseteq_{\text{tree}} T_i' \text{ for all } i.
\]
\end{defn}
With the leaf $\epsilon$ in the role of the undefined tree $\Omega$, this is the order of Courcelle \cite[Section~2.4]{courcelle1983trees} restricted to finite trees.

\begin{prop}[Properties of Tree Order-by-Composition]\label{prop:tree-order-properties}
The relation $\sqsubseteq_{\text{tree}}$ is a partial order on $\mathcal{T}_\Sigma$ with least element $\epsilon$, and:
\begin{enumerate}
    \item $\cost$ and $\depth$ are monotone along $\sqsubseteq_{\text{tree}}$;
    \item $T \sqsubseteq_{\text{tree}} T \otimes T'$ for all $T, T'$;
    \item if $\depth(T) \le d$ then $T \sqsubseteq_{\text{tree}} F_d$, where $F_d$ is the full $b$-ary tree of depth $d$; consequently
    \[
    \cost(T) \le \cost(F_d) = \sum_{k=0}^d b^k = \frac{b^{d+1}-1}{b-1}.
    \]
\end{enumerate}
\end{prop}

\begin{proof}
Reflexivity, transitivity and antisymmetry follow by structural induction from the inductive characterization. (1) A leaf substitution replaces single nodes by non-empty trees. (2) Grafting is a particular leaf substitution. (3) By induction on $T$: a leaf is below every tree; a node of depth at most $d \ge 1$ has children of depth at most $d-1$, each below $F_{d-1}$ by induction. The cost bound then follows from (1).
\end{proof}

\begin{defn}[Uniform Tree Exponentiation]\label{def:tree-exponentiation}
Let $B \in \mathcal{T}_\Sigma$ be a tree representing one level of recursive unfolding. The \textit{uniform tree power} is defined by
\[
B^{\otimes 0} \coloneqq \epsilon, \qquad B^{\otimes (d+1)} \coloneqq B \otimes B^{\otimes d}.
\]
\end{defn}

\begin{prop}[Structural Complexity of Tree Exponentiation]\label{prop:tree-power-complexity}
Let $B$ have $L \coloneqq \leaves(B) \ge 2$ leaves and $I \coloneqq \cost(B) - L$ internal nodes. For every $d \in \mathbb{N}$:
\begin{enumerate}
    \item \textbf{Leaf frontier growth:} $\leaves\left(B^{\otimes d}\right) = L^d$.
    \item \textbf{Cumulative work:} $\cost\left(B^{\otimes d}\right) = I \cdot \dfrac{L^d - 1}{L - 1} + L^d$. In particular, for the elementary branching step $B = g(\epsilon, \dots, \epsilon)$, where $I = 1$ and $L = b$,
    \[
    \cost\left(B^{\otimes d}\right) = \sum_{k=0}^{d} b^k = \frac{b^{d+1} - 1}{b - 1} \in \mathcal{O}(b^d),
    \]
    and $B^{\otimes d}$ is the full tree $F_d$.
    \item \textbf{Prefix chain:} $B^{\otimes 0} \sqsubseteq_{\text{tree}} B^{\otimes 1} \sqsubseteq_{\text{tree}} \dots \sqsubseteq_{\text{tree}} B^{\otimes d}$.
\end{enumerate}
\end{prop}
\begin{proof}
Grafting satisfies $\leaves(T_1 \otimes T_2) = \leaves(T_1) \cdot \leaves(T_2)$ and the \emph{grafting size identity}
\[
\cost(T_1 \otimes T_2) = \left(\cost(T_1) - \leaves(T_1)\right) + \leaves(T_1) \cdot \cost(T_2),
\]
both by structural induction on $T_1$ (\lean{graft\_\allowbreak leafCount}, \lean{graft\_\allowbreak size}). (1) follows by induction on $d$. For (2), the identity gives $\cost(B^{\otimes (d+1)}) = I + L \cdot \cost(B^{\otimes d})$ with $\cost(B^{\otimes 0}) = 1$, whose solution is the stated closed form. For (3), associativity gives $B^{\otimes (d+1)} = B^{\otimes d} \otimes B$, and Proposition~\ref{prop:tree-order-properties}(2) applies.
\end{proof}

\section{Loops and Recursion as Optimization Problems}
\label{sec:optimization}

This section states, for each structured control construct, the optimization problem that characterizes its execution. In each case the statement is an equivalence with the standard big-step (natural) semantics of the construct \cite{winskel1993formal}.

\subsection{While loops}

\begin{defn*}
In imperative programming, a while loop can be written in the generic form

\begin{lstlisting}
while (cond(x)) {
    x = body(x);
}
\end{lstlisting}
which corresponds, in the lambda calculus, to the fixed-point definition \cite{manna1974mathematical}
\[
\text{while} \coloneqq \text{Y}\,\left(\lambda w.\lambda x.\, \text{cond}(x)\, (w\,(\text{body}(x)))\, x\right).
\]
\end{defn*}

\begin{prop}[Loop Execution as Trace Minimization]
\label{prop:min_x}
Let $P : S \to \mathbb{B}$ be the valuation of $\text{cond}$, let $\tau$ be the path selector of $\text{body}$ with induced partial state update $f : S \rightharpoonup S$, and let $s_0 \in S$. The loop terminates from $s_0$ in the state $\bar{s}$ if and only if
\[
\bar{s} = f^{\bar{n}}(s_0), \qquad \bar{n} = \min \left\{ n \in \mathbb{N} \;\middle|\; f^n(s_0) \text{ exits} \right\},
\]
where all iterates $f^k(s_0)$, $k \le \bar{n}$, are defined. The pair $(\bar{n}, \bar{s})$ is unique, and the executed trace $\bar{t} = \rho_{\bar{n}}(s_0)$ is the $\sqsubseteq$-least element of $\{\rho_n(s_0) \mid f^n(s_0) \text{ exits}\}$. For a straight-line body with trace $t$, $\bar{t} = t^{\bar{n}}$ and $\bar{s} = \sem{t^{\bar{n}}}(s_0)$.
\end{prop}

\begin{proof}
The proof of Theorem~\ref{thm:extensional-equivalence} uses only that the body acts on states as the strict extension of a partial function; it applies unchanged to $f$ (\lean{whileBigStep\_\allowbreak iff\_\allowbreak minimalExit}, \lean{MinimalExit.\allowbreak unique}). Minimality of the trace follows from $\rho_n(s_0) \sqsubseteq \rho_m(s_0)$ for $n \le m$.
\end{proof}

Thus executing a loop and solving the minimization problem for $\bar{n}$ are the same thing; a loop is solved, rather than executed, whenever a closed form of $n \mapsto f^n(s_0)$ is available.

\subsection{Linear recursion}

\begin{defn*}
A linear recursion has the generic form
\begin{lstlisting}
function rlin(x) {
    if (cond(x))
        return back(x, rlin(fwd(x)));
    return stop(x);
}
\end{lstlisting}
where \texttt{back} may read the caller's frame \texttt{x}. We call the recursion \textit{frame-independent} if \texttt{back} depends only on the returned value, written $\text{back}(r)$.
\end{defn*}

\begin{prop}[Linear Recursion as Dual-Layer Trace Optimization]
\label{prop:linrec}
Let $P$ be the valuation of $\text{cond}$, and let $\text{fwd} : S \to S$, $\text{stop} : S \to R$ and $\text{back} : S \times R \to R$ be the valuations of the corresponding program fragments. The call $\text{rlin}(s_0)$ returns $r \in R$ if and only if
\[
r = \text{back}\bigl(s_0, \text{back}\bigl(s_1, \dots \text{back}\bigl(s_{\bar{n}-1}, \text{stop}(s_{\bar{n}})\bigr) \dots \bigr)\bigr), \qquad s_k \coloneqq \text{fwd}^k(s_0),
\]
where the recursion depth $\bar{n}$ is the minimal solution of
\[
\bar{n} = \min \left\{ n \in \mathbb{N} \;\middle|\; \neg P\left(\text{fwd}^n(s_0)\right) \right\}.
\]
In the frame-independent case this is the closed form $r = \left(\text{back}^{\bar{n}} \circ \text{stop} \circ \text{fwd}^{\bar{n}}\right)(s_0)$. If $t_{\text{fwd}}, t_{\text{stop}}, t_{\text{back}} \in \Sigma^*$ are straight-line traces of the three fragments (the guard evaluation being accounted for in $t_{\text{fwd}}$ and $t_{\text{stop}}$), the executed trace and its cost are
\[
\bar{t} = t_{\text{fwd}}^{\bar{n}} \cdot t_{\text{stop}} \cdot t_{\text{back}}^{\bar{n}}, \qquad \cost(\bar{t}) = \bar{n} \cdot \left(\cost(t_{\text{fwd}}) + \cost(t_{\text{back}})\right) + \cost(t_{\text{stop}}).
\]
\end{prop}

\begin{proof}
($\Rightarrow$) By induction on the big-step derivation of the call: a base-case call has depth $0$; a recursive call on $s$ with $P(s)$ has depth one more than the call on $\text{fwd}(s)$, and its result is $\text{back}(s, \cdot)$ applied to the inner result. ($\Leftarrow$) By induction on $\bar{n}$. The descending phase is the tail-recursive loop of Proposition~\ref{prop:min_x} with body $\text{fwd}$, which fixes $\bar{n}$; the ascending phase applies $\text{back}$ once per pending frame. The cost identity is additivity of $\cost$ (\lean{framedBigStep\_\allowbreak iff}, \lean{linRec\_\allowbreak iff}, \lean{linear\_\allowbreak recursion\_\allowbreak cost}).
\end{proof}

\begin{rem*}
The tail component of a linear recursion can be isolated in the canonical form
\begin{lstlisting}
function trlin(x) {
  if (cond(x))
    return trlin(fwd(x));
  return x;
}
\end{lstlisting}
whose result $\bar{x} = \text{fwd}^{\bar{n}}(s_0)$ is the first point of the orbit of $s_0$ that falsifies $\text{cond}$.
\end{rem*}

\subsection{Branching recursion}

\begin{defn*}
A non-linear (branching) recursive function with branching factor $b$ has the generic form
\begin{lstlisting}
function rtree(x) {
  if (cond(x))
    return combine(rtree(fwd_1(x)), ..., rtree(fwd_b(x)));
  return stop(x);
}
\end{lstlisting}
\end{defn*}

Positions in a computation tree are addressed by paths $\pi \in \{1, \dots, b\}^*$; the state at position $\pi = i_1 i_2 \cdots i_k$ is $s_\pi \coloneqq (\text{fwd}_{i_k} \circ \dots \circ \text{fwd}_{i_1})(s_0)$. A tree $T$ is \textit{feasible} for $s_0$ if every leaf position $\pi$ of $T$ satisfies $\neg P(s_\pi)$. The \textit{execution tree} of the call $\text{rtree}(s_0)$ is the tree whose internal positions are exactly the calls with $P(s_\pi)$ and whose leaves are the calls with $\neg P(s_\pi)$; it exists if and only if the call terminates.

\begin{prop}[Non-Linear Recursion as Tree Optimization]\label{prop:tree-optimization}
Assume the call $\text{rtree}(s_0)$ terminates. Its execution tree $\bar{T}$ is the least feasible tree:
\[
\bar{T} = \min\nolimits_{\sqsubseteq_{\text{tree}}} \left\{ T \in \mathcal{T}_\Sigma \;\middle|\; \forall\, \pi \in \text{Leaves}(T), \; \neg P\left(s_\pi\right) \right\}.
\]
In particular $\bar{T}$ is unique and has minimal cost and minimal depth among feasible trees. The returned value is the bottom-up catamorphism over $\bar{T}$ that evaluates $\text{stop}(s_\pi)$ at leaves and $\text{combine}$ at internal nodes.
\end{prop}
\begin{proof}
$\bar{T}$ is feasible by definition. Let $T$ be feasible; we show $\bar{T} \sqsubseteq_{\text{tree}} T$ by induction on $\bar{T}$, simultaneously for all initial states. If $\bar{T}$ is a leaf the claim holds since $\epsilon$ is least. Otherwise $P(s_0)$ holds, so $T$ cannot be a leaf (its only leaf would violate feasibility); hence $T$ is a node whose $i$-th subtree is feasible for $\text{fwd}_i(s_0)$, and the induction hypothesis applies to each subtree. Uniqueness follows from antisymmetry, minimality of cost and depth from Proposition~\ref{prop:tree-order-properties}(1), and the statement on the value by induction on the big-step derivation (\lean{exec\_\allowbreak least}, \lean{exec\_\allowbreak unique}, \lean{treeBigStep\_\allowbreak iff}).
\end{proof}

Note that a feasible tree may expand a position where $P$ fails; such trees are strictly above $\bar{T}$, exactly as $t^n$ for $n > \bar{n}$ may exit again without being the executed trace.

\section{Application to Imperative Programming}
\label{sec:application}

We now describe how an imperative program is translated into a constrained optimization problem: first exactly (Section~\ref{subsec:equation-generation}), then in the relaxed form that yields cost bounds (Section~\ref{subsec:asymptotic-relaxation}).

\subsection{Equation generation}
\label{subsec:equation-generation}

The exact translation is a symbolic evaluation in the style of operational semantics \cite{winskel1993formal}: instead of executing the code, it emits unknowns and equations. We describe it at the level of detail needed to follow the examples of Section~\ref{sec:examples}; its justification is Section~\ref{sec:optimization}, and it is not part of the mechanized development.

A symbolic store $\sigma$ maps each program variable $L$ to a family of unknowns $L_0, L_1, \dots$, one per assignment: an assignment $L \coloneqq E$ creates the next unknown of $L$ and emits the equation binding it to the value of $E$, and a read of $L$ evaluates to its current unknown. For each loop three symbolic stores are used: $\sigma_{init}$ (on entry, after the first successful guard), $\sigma_{loop}$ (after $n$ iterations, at the last successful guard) and $\sigma_{last}$ (at the failing guard). They separate the unknowns constrained by the entry condition, by the iterated body and by the exit condition, and they are linked by binding equations: the initial unknown of a variable in a store equals its current unknown in the previous one. The loop itself is translated following Proposition~\ref{prop:min_x}:

\[
\frac{\begin{array}{c}
\left\langle \underset{n\in\mathbb{N}}{\min\,}\mathit{body}^{\circ n}(E),\sigma_{loop}\right\rangle \rightarrow\sigma_{last}\\
\left\langle \mathit{cond}(E),\sigma_{init}\right\rangle \rightarrow \mathit{true}\\
\left\langle \mathit{cond}(E),\sigma_{loop}\right\rangle \rightarrow \mathit{true}\\
\left\langle \mathit{cond}(E),\sigma_{last}\right\rangle \rightarrow \mathit{false}
\end{array}}{\left\langle \mathit{while}(\mathit{cond}(E),\mathit{body}(E)),\sigma\right\rangle \rightarrow\sigma_{last}}
\]

The constraints on $\sigma_{loop}$ and $\sigma_{last}$ express that the guard holds just before the last iteration and fails just after it; together with monotonicity of the guard along the orbit, when available, they determine the minimum. If the guard fails at the initial state the loop is bypassed and the store is unchanged. If no $n \in \mathbb{N}$ satisfies the exit constraint, the system is infeasible and the loop does not terminate (Theorem~\ref{thm:extensional-equivalence}). The term $\mathit{body}^{\circ n}$ can be eliminated whenever the $n$-fold iterate has a closed form, for instance for translations $x \mapsto x + c$, whose iterate is $x \mapsto x + n \cdot c$.

Recursion is translated in the same way. For linear recursion the system consists of the loop constraints for the descending phase, which determine the depth $\bar{n}$, followed by the equations of the ascending phase (Proposition~\ref{prop:linrec}). For branching recursion the unknowns and equations are indexed by the positions $\pi$ of the least feasible tree (Proposition~\ref{prop:tree-optimization}): the guard holds at internal positions and fails at leaves, and values are combined bottom-up. Section~\ref{sec:examples} shows one system of each kind.

\begin{rem*}[Decidability Boundaries]
The generated systems are as hard as the programs they come from. If every update is a translation $x \coloneqq x + c$ and guards are conjunctions of linear inequalities, the iterates are linear in $n$ and the minimization problem is a formula of Presburger arithmetic, hence decidable; this is the setting of loop acceleration \cite{bardin2008fast}, and the examples of Section~\ref{sec:examples} belong to this class. With multiplication between variables, deciding whether the exit constraint has a solution $n$ subsumes the solvability of Diophantine equations, which is undecidable by Matiyasevich's Theorem \cite{matiyasevich1970enumerable}. No complete procedure exists, which motivates the relaxation below.
\end{rem*}

\subsection{Relaxation and WCET Envelopes}
\label{subsec:asymptotic-relaxation}

The exact problem delivers the terminal state $\bar{s}$ and the iteration count $\bar{n}$ together, but it is solvable only on restricted fragments. Moreover, real programs read inputs known only up to a range and branch on data, so that a loop of $n$ iterations may follow up to $2^n$ distinct paths.

In hard real-time systems the terminal state is often not needed. The verification objective is a bound on resource consumption:
\[
\text{WCET}(s_0) \le \text{Deadline}.
\]
CrossCheck therefore relaxes the extensional equations into \textit{cost envelopes}: the state is projected onto a scalar cost through a cost norm $\|\cdot\|$ (Definition~\ref{def:subadditive-norm}), branch alternatives are collapsed with the tropical $(\max, +)$ semiring, and iteration counts are bounded by ranking potentials. Every bound in this subsection is a sound upper bound on the exact cost, not an asymptotic statement.

\subsubsection{Tropical $(\max, +)$ Semiring on Control Flow}
\label{subsubsec:tropical-semiring}

Control-flow alternatives are composed in the idempotent tropical semiring \cite{baccelli1992synchronization}
$\mathbb{N}_{\max, +} \coloneqq (\mathbb{N} \cup \{-\infty\}, \oplus, \otimes)$, where
\[
a \oplus b \coloneqq \max(a, b), \qquad a \otimes b \coloneqq a + b,
\]
with additive identity $-\infty$ and multiplicative identity $0$. Sequencing is the tropical product and branching the tropical sum.

\begin{prop}[Tropical Branch Collapsing]\label{prop:tropical-collapse}
Let a loop body consist of $k$ sequential conditional blocks, the $j$-th with guard trace $t_{c_j}$ and branch traces $t_b$, $b \in \text{Branches}_j$, and let $\tau$ be its path selector. Then for every state $s$,
\[
\|\tau(s)\| \;\le\; W \coloneqq \bigotimes_{j=1}^k \Bigl( \|t_{c_j}\| \otimes \bigoplus_{b \in \text{Branches}_j} \|t_b\| \Bigr) = \sum_{j=1}^k \Bigl( \|t_{c_j}\| + \max_{b \in \text{Branches}_j} \|t_b\| \Bigr).
\]
\end{prop}

\begin{proof}
Apply Proposition~\ref{prop:tropical-dominance} to each block (\lean{branch\_\allowbreak supremum\_\allowbreak bound}, \lean{condTrace\_\allowbreak cost\_\allowbreak le}) and subadditivity across blocks.
\end{proof}

A body with up to $2^k$ distinct paths is thus summarized by the single number $W$, computed in time linear in the size of the body. We write $\|t_{\text{wc}}\| \coloneqq W$ for this worst-case step cost. The bound is exact when the branches attaining the maxima are simultaneously feasible, and an over-approximation otherwise (infeasible paths are not excluded).

\subsubsection{Ranking Potentials}
\label{subsubsec:ranking-potentials}

With a uniform bound $W$ on the cost of one iteration, Proposition~\ref{prop:worst-case-step} reduces the cost of the loop to an upper bound on the minimal exiting iteration count $\bar{n}$. Rather than computing iterates, we use a ranking function in the sense of Floyd \cite{floyd1967assigning}, in the quantitative form used for bound analysis \cite{alias2010multi,podelski2004complete}.

\begin{defn}[Ranking Potential]\label{def:ranking-potential}
Let $f : S \to S$ be the state update of a loop body and $P$ its continuation predicate. A function $V : S \to \mathbb{Z}$ is a \textit{ranking potential} with step decrease $\Delta \in \mathbb{Z}$, $\Delta > 0$, if for all states $s \in S$ with $P(s) = \text{true}$:
\begin{enumerate}
    \item \textbf{Descent:} $V\left(f(s)\right) \le V(s) - \Delta$;
    \item \textbf{Positivity:} $V(s) > 0$.
\end{enumerate}
\end{defn}

Both conditions are required only on continuing states; in practice they are established on an inductive invariant containing the reachable states. A different exit threshold $V(s) > V_{\text{exit}}$ is covered by replacing $V$ with $V - V_{\text{exit}}$.

\begin{thm}[Lyapunov Cost Dominance Bound]\label{thm:lyapunov-bound}
Let $V$ be a ranking potential with step decrease $\Delta$ for a loop with update $f$, continuation predicate $P$ and path selector $\tau$. For every initial state $s_0 \in S$ the loop terminates, and its minimal exiting iteration count satisfies
\[
\bar{n} \;\le\; \bar{N}(s_0) \coloneqq \max\left(0, \left\lceil \frac{V(s_0)}{\Delta} \right\rceil\right).
\]
Consequently, if $\|\tau(s)\| \le \|t_{\text{wc}}\|$ for all $s$, the executed trace satisfies
\[
\left\|\rho_{\bar{n}}(s_0)\right\| \;\le\; \bar{N}(s_0) \cdot \|t_{\text{wc}}\|.
\]
\end{thm}

\begin{proof}
Write $s_k \coloneqq f^k(s_0)$ and suppose $P(s_j)$ holds for all $j < n$. By induction on $k \le n$, descent gives $V(s_k) \le V(s_0) - k \cdot \Delta$. If $n \ge 1$, positivity at $s_{n-1}$ gives $0 < V(s_{n-1}) \le V(s_0) - (n-1) \cdot \Delta$, hence $n - 1 < V(s_0)/\Delta$ and, $n$ being an integer, $n \le \lceil V(s_0)/\Delta \rceil$. So every run of $n$ consecutive continuing iterations has $n \le \bar{N}(s_0)$. In particular $P$ cannot hold at all of $s_0, \dots, s_{\bar{N}(s_0)}$: some iterate exits, the minimal exiting count $\bar{n}$ exists by Theorem~\ref{thm:minimal-trace}, and applying the bound to $n = \bar{n}$ gives $\bar{n} \le \bar{N}(s_0)$. The cost bound is Proposition~\ref{prop:worst-case-step} (\lean{ranking\_\allowbreak bound}, \lean{ranking\_\allowbreak terminates}, \lean{wcet\_\allowbreak bound}).
\end{proof}

The theorem is stated for integer-valued potentials and a deterministic update. The same argument gives a variant that is closer to what a tool computes (\lean{ranking\_\allowbreak bound\_\allowbreak floor}): let $V$ take values in an ordered field (for instance $\mathbb{Q}$) and let $\to$ be a transition \emph{relation} on $S$, possibly nondeterministic or partial; if $V(s') \le V(s) - 1$ and $V(s) \ge 1$ whenever $s \to s'$, then every run $s_0 \to s_1 \to \dots \to s_n$ has $n \le \lfloor V(s_0) \rfloor$. A rational potential with decrease $\Delta$ is brought to this form by dividing by $\Delta$.

In practice the conditions on $V$ hold only on the states that a run can reach, and the bound on the initial potential depends on what is known about the initial state. The next statement separates the two.

\begin{prop}[Bound under an invariant and a context]
\label{prop:context-bound}
Let $\to$ be a transition relation on $S$, let $V : S \to \mathbb{Q}$, and let $I, C \subseteq S$. Assume that
\begin{enumerate}
    \item $I$ is preserved by steps: if $s \in I$ and $s \to s'$, then $s' \in I$;
    \item $V(s') \le V(s) - 1$ and $V(s) \ge 1$ whenever $s \in I$ and $s \to s'$;
    \item $V(s) \le K$ for every $s \in C$.
\end{enumerate}
Then every run $s_0 \to s_1 \to \dots \to s_n$ with $s_0 \in I \cap C$ satisfies $s_k \in I$ for all $k \le n$, and $n \le \lfloor K \rfloor$.
\end{prop}

\begin{proof}
By induction on $k$ and assumption~(1), $s_k \in I$ for every $k \le n$. The run is therefore a run of the relation $\to_I$ defined by $s \to_I s'$ if and only if $s \in I$ and $s \to s'$, and by assumption~(2) the potential $V$ satisfies the hypotheses of the relational variant above for $\to_I$. Hence $n \le \lfloor V(s_0) \rfloor$, and $\lfloor V(s_0) \rfloor \le \lfloor K \rfloor$ by assumption~(3), since $s_0 \in C$.
\end{proof}

The set $I$ is an inductive invariant of the loop. The set $C$ is a \emph{context}: a constraint on the initial state, for instance on the values of the parameters of a function at its entry. Two consequences are used by the implementation of the companion paper \cite{crosscheckTool}. First, since every state of the run is in $I$, the invariant holds wherever the run stops, in particular when a loop is left by an early exit and not by its guard. Second, the bound is a function of the context: if a cost bound $B$ has been derived for a function under a context $C$, it applies to every call whose arguments satisfy $C$, so that $m$ such calls cost at most $m \cdot B$; a function may thus be given different bounds at different call sites, each under a context that is proved at that site.

Two classical shapes of termination argument give bounds for loops that have no single affine potential; Floyd \cite{floyd1967assigning} already used, besides the number of steps until termination or an upper bound on it, $n$-tuples of non-negative integers ordered lexicographically; lexicographic and multi-dimensional rankings are studied in \cite{alias2010multi,benamram2014ranking}.

\begin{prop}[Lexicographic and logarithmic potentials]
\label{prop:lex-log}
Let $\to$ be a transition relation on $S$ and $I \subseteq S$ a set preserved by steps.
\begin{enumerate}
    \item Let $V_1, V_2 : S \to \mathbb{Q}$ and $K_2 \in \mathbb{Q}$ with $V_1 \ge 0$ and $0 \le V_2 \le K_2$ on $I$. If every step $s \to s'$ with $s \in I$ either satisfies $V_1(s') \le V_1(s) - 1$ and $V_1(s) \ge 1$, or satisfies $V_1(s') \le V_1(s)$, $V_2(s') \le V_2(s) - 1$ and $V_2(s) \ge 1$, then every run from $s_0 \in I$ has at most $\lfloor V_1(s_0) (K_2 + 1) + V_2(s_0) \rfloor$ steps.
    \item Let $V : S \to \mathbb{N}$ and $b \ge 2$ an integer. If every step $s \to s'$ with $s \in I$ satisfies $b \cdot V(s') \le V(s)$ and $V(s) \ge 1$, then every run from $s_0 \in I$ has at most $\lfloor \log_b V(s_0) \rfloor + 1$ steps.
\end{enumerate}
\end{prop}

\begin{proof}
(1) Let $W = V_1 (K_2 + 1) + V_2$. A step of the first kind decreases $V_1 (K_2 + 1)$ by at least $K_2 + 1$ and increases $V_2$ by at most $K_2$, since $V_2(s') \le K_2$ and $V_2(s) \ge 0$; a step of the second kind does not increase $V_1 (K_2 + 1)$ and decreases $V_2$ by at least $1$. In both cases $W(s') \le W(s) - 1$, and $W(s) \ge 1$ because either $V_1(s) \ge 1$ and $V_2(s) \ge 0$, or $V_2(s) \ge 1$ and $V_1(s) \ge 0$. Proposition~\ref{prop:context-bound} applies to $W$. (2) By induction, $b^k V(s_k) \le V(s_0)$ along a run. If the run has $n \ge 1$ steps, the last one starts from $s_{n-1}$ with $V(s_{n-1}) \ge 1$, so $b^{n-1} \le V(s_0)$ and $n - 1 \le \lfloor \log_b V(s_0) \rfloor$.
\end{proof}

The first statement covers loops in which an inner counter is reset when an outer one advances. Its proof shows more than a bound on runs: $W$ satisfies the two premises of the loop rule, decrease by $1$ and $W \ge 1$ before a step (\lean{lex\_\allowbreak potential}), so that with affine $V_1$, $V_2$ and a constant $K_2$ the pair is one affine potential, found by the synthesis of Section~\ref{subsubsec:smt-synthesis} as soon as the bounds of $V_2$ are among the invariants; the second covers halving loops such as binary search, for which an affine potential gives only a linear bound.

\subsubsection{Hierarchical and Recursive Composition}
\label{subsubsec:compositional-wcet}

\paragraph{Nested loops.}
Let the outer body consist of a straight-line part $t_1$ followed by an inner loop with body cost at most $\|t_2\|$. Suppose the inner loop has a ranking potential $V_2$ with decrease $\Delta_2$, and let $M_{\max}$ be an upper bound of $\lceil V_2(s)/\Delta_2 \rceil$ over all states $s$ at which the inner loop can be entered. Then every outer iteration costs at most $\|t_1\| + M_{\max} \cdot \|t_2\|$, and if the outer loop has a ranking potential with bound $N_{\max}$, Proposition~\ref{prop:worst-case-step} yields the product envelope (compare Proposition~\ref{prop:nested-cost})
\[
\left\|\rho_{\bar{n}}(s_0)\right\| \le N_{\max} \cdot \|t_1\| + N_{\max} \cdot M_{\max} \cdot \|t_2\|
\]
(\lean{nested\_\allowbreak wcet\_\allowbreak bound}). This bound is sound but not amortized: when the inner count varies across outer iterations, methods such as \cite{sinn2014complexity,hoffmann2012multivariate} can be asymptotically more precise. Two cases of amortization are covered below (Proposition~\ref{prop:amortised-loop}): when the cost of an outer iteration is at most $a + c\,V$, with $V$ the outer potential before the iteration, as in a triangular nest whose inner loop runs from the outer counter to a fixed limit, or at most $a + c\,(K - V)$, as when it runs from $0$ to the outer counter, the outer loop costs at most the sum of these costs over the iterations, about half the product bound.

\paragraph{Branching recursion.}
For the recursion scheme of Proposition~\ref{prop:tree-optimization}, a \textit{depth ranking potential} is a function $V_{\text{tree}} : S \to \mathbb{N}$ such that, for all $s$ with $P(s) = \text{true}$ and all $i \in \{1, \dots, b\}$,
\[
V_{\text{tree}}\left(\text{fwd}_i(s)\right) < V_{\text{tree}}(s).
\]

\begin{prop}[Branching Recursion Envelope]\label{prop:tree-envelope}
If $V_{\text{tree}}$ is a depth ranking potential and the call $\text{rtree}(s_0)$ has execution tree $\bar{T}$, then with $D_{\max} \coloneqq V_{\text{tree}}(s_0)$,
\begin{gather*}
\depth(\bar{T}) \le D_{\max}, \\
\cost(\bar{T}) \le \cost\left(B^{\otimes D_{\max}}\right) = \frac{b^{D_{\max} + 1} - 1}{b - 1} \in \mathcal{O}(b^{D_{\max}}),
\end{gather*}
where $B$ is the elementary $b$-ary branching step.
\end{prop}

\begin{proof}
By induction on $\bar{T}$, simultaneously for all initial states: a leaf has depth $0$; at an internal node $P(s_0)$ holds, each subtree is the execution tree of $\text{fwd}_i(s_0)$ and has depth at most $V_{\text{tree}}(\text{fwd}_i(s_0)) \le V_{\text{tree}}(s_0) - 1$. The cost bound is Proposition~\ref{prop:tree-order-properties}(3) together with Proposition~\ref{prop:tree-power-complexity}(2) (\lean{exec\_\allowbreak depth\_\allowbreak le}, \lean{exec\_\allowbreak size\_\allowbreak le}).
\end{proof}

The envelope counts calls. If each call performs at most $w$ units of local work, the total cost is at most $w \cdot \cost(\bar{T})$. When the local work depends on the size of the argument, as in typical divide-and-conquer algorithms, sharper bounds require solving the corresponding recurrence, e.g.\ by the Master Theorem \cite{cormen2009introduction}; this refinement is outside the scope of the present paper.

The depth envelope is exponential in $D_{\max}$ even when the recursion is balanced and its tree is small. A second certificate bounds the number of calls directly. It is stated for a general recursive scheme: an activation with argument $x$ performs self-calls with arguments $c_1(x), \dots, c_{k(x)}(x)$, where $k(x) = 0$ for base cases, and the \textit{call tree} of $x$ has a root labelled $x$ whose subtrees are the call trees of the $c_i(x)$.

\begin{prop}[Size Potential]\label{prop:size-potential}
Let $U : S \to \mathbb{Q}$ satisfy, for every argument $x$,
\[
U(x) \;\ge\; 1 + \sum_{i=1}^{k(x)} U\left(c_i(x)\right).
\]
If the call with argument $x$ terminates, its call tree has at most $\lfloor U(x) \rfloor$ nodes. If moreover $k(x) \le b$ for all $x$ and $V : S \to \mathbb{N}$ strictly decreases at every self-call, the call tree has at most $\sum_{j=0}^{V(x)} b^j$ nodes.
\end{prop}

\begin{proof}
By induction on the call tree: its size is $1$ plus the sizes of its subtrees, which by induction are at most $U(c_i(x))$. The second statement is Proposition~\ref{prop:tree-envelope} for trees whose nodes have at most $b$ children (\lean{callTree\_\allowbreak size\_\allowbreak le\_\allowbreak potential}, \lean{callTree\_\allowbreak size\_\allowbreak le\_\allowbreak geom}).
\end{proof}

This is the potential method of amortized analysis \cite{tarjan1985complexity} (whose potential view Tarjan credits to Sleator), as used in automatic resource analysis \cite{hoffmann2012multivariate}, specialized to counting calls. For a function that splits a range $[lo, hi)$ with $lo < hi$ at an interior point and recurses on both halves, $U(lo, hi) = 2(hi - lo) - 1$ is a size potential, so the number of calls is linear in the size of the range.

\subsubsection{SMT-Based Synthesis of Ranking Potentials}
\label{subsubsec:smt-synthesis}

Finding $V$ does not require interactive proof. For a loop over integer variables $\mathbf{x} \in \mathbb{Z}^k$ with affine update $\mathbf{x}' = A\mathbf{x} + \mathbf{b}$ and a guard (strengthened by a known invariant) given as a conjunction of linear inequalities $G\mathbf{x} \le \mathbf{g}$, CrossCheck uses the linear template
\[
V(\mathbf{x}) = \mathbf{w}^T \mathbf{x} + c_0, \qquad \mathbf{w} \in \mathbb{Z}^k, \; c_0 \in \mathbb{Z}.
\]
The conditions of Definition~\ref{def:ranking-potential} become the universally quantified implications
\[
G\mathbf{x} \le \mathbf{g} \implies \mathbf{w}^T (\mathbf{x} - A\mathbf{x} - \mathbf{b}) \ge \Delta, \qquad G\mathbf{x} \le \mathbf{g} \implies \mathbf{w}^T \mathbf{x} + c_0 \ge 1 .
\]
By the affine form of Farkas' Lemma \cite{schrijver1986theory}, each implication between linear inequalities whose premises are satisfiable is equivalent, over the rationals, to the existence of non-negative multipliers satisfying linear equalities (Corollary~7.1h there). The synthesis problem thus becomes a quantifier-free linear arithmetic query in the unknowns $\mathbf{w}$, $c_0$ and the multipliers \cite{colon2001synthesis,podelski2004complete}, which is discharged by an SMT solver such as Z3 \cite{demoura2008z3}. Only the soundness direction of the lemma, from multipliers to the implication, is needed to trust a result; it does not need the premises to be satisfiable and it is elementary (\lean{farkas\_\allowbreak sound\_\allowbreak affine}). The method is sound for integer programs and complete for linear ranking functions over the rationals \cite{benamram2014ranking}; a loop that terminates only by a non-linear or lexicographic argument is not handled by this template.

When the query is satisfiable, the model provides $\mathbf{w}$ and $c_0$, and Theorem~\ref{thm:lyapunov-bound} gives the closed-form certificate
\[
\max\left(0, \left\lceil \frac{\mathbf{w}^T \mathbf{x}_0 + c_0}{\Delta} \right\rceil\right) \cdot \|t_{\text{wc}}\| \;\le\; \text{Deadline},
\]
to be checked for the admissible range of initial values $\mathbf{x}_0$. The certificate is independent of the solver: given $\mathbf{w}$ and $c_0$, the two implications above can be re-checked directly, so the solver need not be trusted. When the query is unsatisfiable, the routine is rejected: the template admits no linear potential over the rationals for the given guard and invariant, and a timing bound cannot be established by this method.

\subsection{A cost logic for a small language}
\label{subsec:cost-logic}

The bounds above concern one loop or one recursion at a time. To state that they compose correctly over whole programs, we fix a small language and a proof system for cost bounds, and prove the proof system sound. Hoare-style logics for running time go back to Nielson's thesis \cite{nielson1984hoare} and journal paper \cite{nielson1987hoare}; Shaw extended Hoare logic with a real-time variable advanced by the timing bound of each statement \cite{shaw1989reasoning}, and Haslbeck and Nipkow \cite{haslbeck2018hoare} formalized three such logics in Isabelle/HOL for the deterministic language IMP (a logic in the style of Nielson, a logic with potentials and a separation logic with time credits), proved each of them sound and complete, and gave verified verification-condition generators. The system below is of this kind, and we do not claim the idea as new; it is weaker in that only soundness is proved. It differs in three respects, which are those needed to connect it to the analysis of this paper. Its atomic actions are relations, so that unknown inputs and partial operations are covered; in \cite{haslbeck2018hoare} a valid triple asserts that the program terminates within the bound, which is the appropriate notion for a deterministic language, whereas here a triple bounds the cost of every execution, including executions that do not terminate. Its loop rule takes as premises exactly the certificate of Proposition~\ref{prop:context-bound}: an invariant, a rational potential and a context. Nielson's logic proves total correctness together with a bound on the computation time, expressed in the initial values of the variables; its loop rule bounds the number of unfoldings by a natural-number variable \cite[rules R-3, R-4]{nielson1987hoare}.

\paragraph{Language.}
Let $S$ be a set of states. \emph{Commands} are given by
\[
c \;::=\; \mathtt{skip} \;\mid\; a \;\mid\; c_1 ; c_2 \;\mid\; \mathtt{if}\ b\ \mathtt{then}\ c_1\ \mathtt{else}\ c_2 \;\mid\; \mathtt{while}\ b\ \mathtt{do}\ c \;\mid\; \mathtt{call}\ p,
\]
where an atomic action $a$ is a relation ${\to_a} \subseteq S \times S$ with a weight $\omega(a) \in \mathbb{N}$, a test $b : S \to \{\mathit{true}, \mathit{false}\}$ has a weight $\omega(b) \in \mathbb{N}$, and $p$ ranges over a set of procedure names; a call has a weight $\omega_{\mathrm{call}} \in \mathbb{N}$, and an environment $\Phi$ assigns to every procedure name a command, its body, which may call any procedure, itself included. Parameter passing is not primitive: arguments and results are passed through the state by atomic actions. We write $c : s \Downarrow_n t$ when $c$, started in $s$, terminates in $t$ with cost $n$; this relation is defined by the usual big-step rules, in which an atomic step $s \to_a t$ costs $\omega(a)$, each evaluation of a test costs its weight, a call costs $\omega_{\mathrm{call}}$ plus the cost of the body, and costs add along the execution. We write $c : s \uparrow n$ when some execution of $c$ from $s$, \emph{terminated or not}, has accumulated cost $n$: this relation contains $c : s \uparrow 0$, every $n$ with $c : s \Downarrow_n t$, and is closed under the rules that let an execution stop inside the first or the second component of a sequence, inside a branch, inside or after any iteration of a loop body, and inside the body of a called procedure. A program that diverges has prefixes of every length; bounding $\uparrow$ therefore bounds its cost as well.

\begin{defn}[Valid cost triple]
\label{def:valid-triple}
For $P, Q \subseteq S$, a command $c$ and $B \in \mathbb{N}$, the triple $\{P\}\ c\ \{Q\}\!\downarrow\! B$ is \emph{valid} if for every $s \in P$: every $n$ with $c : s \uparrow n$ satisfies $n \le B$, and every $t$ with $c : s \Downarrow_n t$ is in $Q$.
\end{defn}

\begin{defn}[Derivable cost triple]
\label{def:derivable-triple}
Let $\Gamma$ be a set of cost triples, the \emph{assumed} triples. The triples \emph{derivable from} $\Gamma$ are generated by the following rules, where $b$ also denotes the set of states in which the test is true and $\neg b$ its complement.
\begin{enumerate}
    \item $\{P\}\ \mathtt{skip}\ \{P\}\!\downarrow\! 0$.
    \item $\{P\}\ a\ \{Q\}\!\downarrow\! \omega(a)$, if $t \in Q$ whenever $s \in P$ and $s \to_a t$.
    \item If $\{P\}\ c_1\ \{R\}\!\downarrow\! B_1$ and $\{R\}\ c_2\ \{Q\}\!\downarrow\! B_2$, then $\{P\}\ c_1 ; c_2\ \{Q\}\!\downarrow\! B_1 + B_2$.
    \item If $\{P \cap b\}\ c_1\ \{Q\}\!\downarrow\! B_1$ and $\{P \cap \neg b\}\ c_2\ \{Q\}\!\downarrow\! B_2$, then
    \[
    \{P\}\ \mathtt{if}\ b\ \mathtt{then}\ c_1\ \mathtt{else}\ c_2\ \{Q\}\!\downarrow\! \omega(b) + \max(B_1, B_2).
    \]
    \item (Loop.) Let $I, C \subseteq S$, $V : S \to \mathbb{Q}$ and $K \in \mathbb{Q}$. If, for every $v \in \mathbb{Q}$, $\{I \cap b \cap \{V = v\}\}\ c\ \{I \cap \{V \le v - 1\}\}\!\downarrow\! W$, if $V(s) \ge 1$ for every $s \in I \cap b$, and if $V(s) \le K$ for every $s \in C$, then
    \[
    \{I \cap C\}\ \mathtt{while}\ b\ \mathtt{do}\ c\ \{I \cap \neg b\}\!\downarrow\! \lfloor K \rfloor \cdot (\omega(b) + W) + \omega(b),
    \]
    where $\lfloor K \rfloor$ is read as $0$ when $K < 0$.
    \item (Consequence.) If $P' \subseteq P$, $\{P\}\ c\ \{Q\}\!\downarrow\! B$, $B \le B'$ and $Q \subseteq Q'$, then $\{P'\}\ c\ \{Q'\}\!\downarrow\! B'$.
    \item (Dead code.) $\{\emptyset\}\ c\ \{Q\}\!\downarrow\! 0$.
    \item (Assumption.) Every triple of $\Gamma$.
    \item (Call.) Let $(\{P_k\}\ \mathtt{call}\ p_k\ \{Q_k\}\!\downarrow\! \omega_{\mathrm{call}} + B_k)_{k \in \kappa}$ be a family of triples about calls, with a rank $r(k) \in \mathbb{N}$ for each index. If, for every $k$, the triple $\{P_k\}\ \Phi(p_k)\ \{Q_k\}\!\downarrow\! B_k$ about the \emph{body} is derivable from $\Gamma$ together with the triples of the family whose rank is smaller than $r(k)$, then every triple of the family is derivable from $\Gamma$.
\end{enumerate}
\end{defn}

Rules (3) and (4) are the sum and maximum rules of Section~\ref{subsubsec:tropical-semiring}; rule (5) combines Proposition~\ref{prop:context-bound} with the loop cost formula; rule (6) is the rule by which a bound derived under a context is used wherever the context holds. Rule (8) is how code that is not analyzed enters a derivation: the contract of a library function, or a cost stated by an annotation, is an assumed triple about the command that stands for it. Rule (9) is the usual rule for recursive procedures \cite{nipkow2002hoare}, with the well-founded argument made explicit by the rank: typically the family is indexed by the value $k$ of a measure of the arguments, $P_k$ states that the measure is at most $k$, and $B_k$ is the bound for that value; for a procedure that is not recursive, the family has one element and no assumption is used.

\begin{thm}[Soundness of the cost logic]
\label{thm:cost-logic-soundness}
If every triple of $\Gamma$ is valid, every cost triple derivable from $\Gamma$ is valid. In particular, every triple derivable from no assumption is valid.
\end{thm}

\begin{proof}
By induction on the derivation, for all sets $\Gamma$ of valid triples simultaneously. Rules (7) and (8) are immediate. For rules (1) and (2) the executions and prefixes of the command are immediate from the definitions. For rule (3), a terminating execution of $c_1 ; c_2$ from $s \in P$ is an execution $c_1 : s \Downarrow_{n_1} u$ followed by $c_2 : u \Downarrow_{n_2} t$; by the induction hypotheses $n_1 \le B_1$, $u \in R$, $n_2 \le B_2$ and $t \in Q$. A prefix is either a prefix of $c_1$, of cost at most $B_1$, or an execution $c_1 : s \Downarrow_{n_1} u$ followed by a prefix of $c_2$ from $u \in R$, of cost at most $n_1 + B_2$. Rule (4) is similar, the test being paid once and one branch being executed. Rule (6) is immediate.

For rule (5), write $w = \omega(b)$ and let $L$ be the loop. We show, for every $m \in \mathbb{N}$ and every $s \in I$ with $V(s) < m + 1$: (i) if $L : s \Downarrow_n t$ then $n \le m (w + W) + w$ and $t \in I \cap \neg b$; (ii) if $L : s \uparrow n$ then $n \le m (w + W) + w$. Both are proved by induction on the derivation of the execution, for all $m$ simultaneously. If the test is false in $s$, the execution is the single test, of cost $w$, and $t = s$. If it is true, then $V(s) \ge 1$, hence $m \ge 1$; the body executes as $c : s \Downarrow_{n_1} u$, and by the induction hypothesis on the derivation of the triple, applied with $v = V(s)$, we get $n_1 \le W$, $u \in I$ and $V(u) \le V(s) - 1 < m$. The rest of the execution is an execution (for (i)) or a prefix (for (ii)) of $L$ from $u$, to which the induction hypothesis applies with $m - 1$; the total cost is at most $w + W + (m - 1)(w + W) + w = m (w + W) + w$. In case (ii) the execution may also stop inside the body of the current iteration, after a prefix of $c$ of cost at most $W$, for a total of at most $w + W \le m (w + W) + w$. Finally, for $s \in I \cap C$ we have $V(s) \le K < \lfloor K \rfloor + 1$, so the claim applies with $m = \lfloor K \rfloor$.

For rule (9), we show by strong induction on $n$ that every triple of the family of rank $n$ is valid. Let $r(k) = n$. The triples of the family of smaller rank are valid by the induction hypothesis on $n$, and those of $\Gamma$ by assumption; by the induction hypothesis on the derivation, the triple $\{P_k\}\ \Phi(p_k)\ \{Q_k\}\!\downarrow\! B_k$ is therefore valid. An execution of $\mathtt{call}\ p_k$, terminated or not, is the payment of $\omega_{\mathrm{call}}$ followed by an execution of the body, so the triple about the call is valid with the bound $\omega_{\mathrm{call}} + B_k$.
\end{proof}

The loop rule pays every iteration the largest cost $W$ of the body. When the cost of an iteration depends on the potential before it, a sharper bound is the sum over the iterations; the next statement is a derived rule, stated on executions. In a Hoare logic for running time, this is already the while rule of Nielson \cite{nielson1987hoare}: the number of unfoldings of the loop is bounded by a variable $z$, and the time of the loop is a solution of a recurrence relation in the unfoldings; Nielson's Section~5 derives in this way the summing series $T(i) = (i - 1) + T(i - 1)$ of bubble sort, a triangular nest. Summing the iteration counts of a non-rectangular loop nest is also classical in WCET analysis: Healy et al.\ \cite{healy2000supporting} formulate the number of iterations of an inner loop whose bounds depend on an outer counter as a summation, following Sakellariou's summations for load balancing, and use it for tighter timing predictions; in cost analysis, Albert, Genaim and Masud \cite{albert2011more} bound the cost of the successive iterations of a loop by an arithmetic (or geometric) sequence, increasing or decreasing, and sum it, instead of multiplying the number of iterations by the largest cost. The rule below is this summation as a derived rule of the cost logic of this paper, with a case of the checker of Section~\ref{subsec:checker} and a Lean proof, in two forms: a cost that grows with the potential (part 1) and, more generally, any cost bounded through a monotone function of the potential (part 2), which covers a cost that grows as the potential falls.

\begin{prop}[Amortised loop rule]
\label{prop:amortised-loop}
Let $I \subseteq S$, a test $b$ of weight $w = \omega(b)$, a body $c$, an integer potential $V : S \to \mathbb{Z}$, an integer $d \ge 1$ and functions $W, F : \mathbb{Z} \to \mathbb{N}$ be such that, for every $s \in I \cap b$: every prefix of $c$ from $s$ costs at most $W(V(s))$, every terminating execution $c : s \Downarrow_n t$ ends in $t \in I$ with $V(t) + d \le V(s)$, and $V(s) \ge d$. (1) If $F(X) \ge w$ for every $X$, and $w + W(v) + F(X - d) \le F(X)$ whenever $d \le v \le X$, then, for every $K \in \mathbb{Z}$,
\[
\{I \cap \{V \le K\}\}\ \mathtt{while}\ b\ \mathtt{do}\ c\ \{I \cap \neg b\}\!\downarrow\! F(K).
\]
For $W(v) = a + c' \max(0, v)$ with $a, c' \in \mathbb{N}$, the function
\[
F(X) = (N + 1)\,w + N a + c' \left(N X - d\,\frac{N (N - 1)}{2}\right), \qquad N = \max(0, \lfloor X / d \rfloor),
\]
satisfies both conditions; $F(K)$ is the sum, over at most $N$ iterations starting with $V \le K, K - d, K - 2d, \dots$, of $w + a + c' V$, plus the last test.

(2) The same triple holds if, instead of the last two conditions, $F$ is monotone, $F(X) \ge w$ for every $X$, and $w + W(v) + F(v - d) \le F(v)$ whenever $v \ge d$. For $W(v) = a + c' \max(0, K - v)$ (an iteration costs more when the potential has fallen further), the function $F(X) = (N + 1)\,w + \sum_{m=1}^{N} \big(a + c' \max(0, K - m d)\big)$, with $N = \max(0, \lfloor X / d \rfloor)$, satisfies these conditions, and
\[
F(K) = (N + 1)\,w + N a + c' \left(N K - d\,\frac{N (N + 1)}{2}\right), \qquad N = \max(0, \lfloor K / d \rfloor).
\]
\end{prop}

\begin{proof}
(1) We show, by induction on the derivation of an execution (terminated or not) of the loop from $s \in I$, that for every $X \ge V(s)$ its cost is at most $F(X)$, and that a terminated one ends in $I \cap \neg b$. If the test is false, the cost is $w \le F(X)$. If it is true, then $d \le V(s) \le X$; a prefix that stops inside the body costs at most $w + W(V(s)) \le F(X)$; otherwise the body ends in $u \in I$ with $V(u) \le V(s) - d \le X - d$, and the rest of the execution costs at most $F(X - d)$ by the induction hypothesis, for a total of at most $w + W(V(s)) + F(X - d) \le F(X)$. For the closed form, all quantities are integers once doubled, and $N(N-1)$ is even. If $X \ge d$, then $\lfloor (X - d)/d \rfloor = N - 1$, and expanding gives $F(X) - F(X - d) = w + a + c' X \ge w + W(v)$ for $v \le X$. Finally $F(X) \ge w$: if $N = 0$ this is equality; otherwise $N d \le X$, so $2 N X - d N (N - 1) \ge 2 N^2 d - d N^2 + d N \ge 0$.

(2) The same induction shows that an execution from $s \in I$ costs at most $F(V(s))$, which is at most $F(K)$ by monotonicity when $V(s) \le K$: if the test is true, the body ends in $u \in I$ with $V(u) \le V(s) - d$, the rest costs at most $F(V(u)) \le F(V(s) - d)$, and the total is at most $w + W(V(s)) + F(V(s) - d) \le F(V(s))$. In the example, $F$ is monotone because $N$ is monotone in $X$ and the terms are natural numbers, and $F(X) \ge w$. For $v \ge d$, $N = \lfloor v/d \rfloor \ge 1$ and $\lfloor (v - d)/d \rfloor = N - 1$, so $F(v) - F(v - d) = w + a + c' \max(0, K - N d) \ge w + W(v)$, since $N d \le v$. For $X = K$ and $m \le N$, $K - m d \ge 0$, and the sum is $N a + c' (N K - d N (N + 1)/2)$. (Part (1) bounds the rest of the run by the bound $X$ of the potential, part (2) by its value; with a decreasing $W$ the hypothesis of part (1) would require $W(d) \le W(X)$.)
\end{proof}

A bound on all execution prefixes also excludes divergence. An \emph{infinite execution} of a command is defined coinductively, as in \cite{leroy2009coinductive}, by the rules of $\Downarrow$ read as non-terminating: a sequence runs forever in its first component, or terminates it and runs forever in the second; a conditional runs forever in the branch it takes; a loop whose test holds runs forever in its body, or completes the body and runs forever in the rest of the loop; a call runs forever in the body of the procedure.

\begin{prop}[No infinite execution]
\label{prop:no-divergence}
Assume that every loop test and every call has a positive weight. Then a command with an infinite execution from $s$ has, for every $N$, an execution prefix from $s$ of cost at least $N$. Consequently, if $\{P\}\ c\ \{Q\}\!\downarrow\! B$ is valid, in particular if it is derivable from no assumption, $c$ has no infinite execution from any state of $P$.
\end{prop}
\begin{proof}
By induction on $N$ and, inside it, on the command. A loop that completes its body and continues, and a call, pay a positive weight before the rest of the infinite execution, to which the induction hypothesis on $N$ applies; a sequence, a conditional and a loop that runs forever in its body continue with a smaller command, to which the inner induction hypothesis applies. A prefix of cost $B + 1$ contradicts validity, and derivable triples are valid by Theorem~\ref{thm:cost-logic-soundness} (\lean{div\_\allowbreak prefixes\_\allowbreak unbounded}, \lean{valid\_\allowbreak no\_\allowbreak div}, \lean{derivable\_\allowbreak no\_\allowbreak div}; the example \lean{spin\_\allowbreak no\_\allowbreak bound} shows that the hypotheses are satisfiable).
\end{proof}

\paragraph{Symbolic bounds and function summaries.}
A bound may depend on the initial state. For $B : S \to \mathbb{N}$, we say that $\{P\}\ c\ \{Q\}\!\downarrow\! B$ holds \emph{symbolically} if, for every $k \in \mathbb{N}$, the triple $\{P \cap \{B \le k\}\}\ c\ \{Q\}\!\downarrow\! k$ is valid; equivalently, for every $s \in P$, every prefix of $c$ from $s$ costs at most $B(s)$ and every terminating execution ends in $Q$. Symbolic bounds are thus expressed with numeric triples, and they are the form in which the bound of a function can be stated once and used at every call:

\begin{prop}[Call rule for summaries]
\label{prop:summary-call}
Let $\{P\}\ c\ \{Q\}\!\downarrow\! B$ hold symbolically, and let $P' \subseteq S$ and $k \in \mathbb{N}$ be such that $s \in P$ and $B(s) \le k$ for every $s \in P'$. Then $\{P'\}\ c\ \{Q\}\!\downarrow\! k$ is valid. Moreover, if $\{R\}\ c_2\ \{Q\}\!\downarrow\! B_2$ holds symbolically, every terminating execution of $c$ from $P$ ends in $R$, and $B_2(u) \le B_2'(s)$ whenever $s \in P$ and $c : s \Downarrow_n u$, then $\{P\}\ c ; c_2\ \{Q\}\!\downarrow\! B + B_2'$ holds symbolically.
\end{prop}

\begin{proof}
The first statement is rule (6) applied to the triple for the budget $k$. The second follows from the state-by-state reading of symbolic bounds, as in the case of rule (3) above, with $B_2(u) \le B_2'(s)$ used to express the cost of $c_2$ in terms of the initial state.
\end{proof}

The second statement is the composition rule of Nielson's logic \cite[rule R-3]{nielson1987hoare}, in which the time of $c_2$ is related to the initial values of the variables through the relation that $c_1$ establishes between its initial and final states.

A calling context is the special case in which $B$ is constant on $P$. The bounds of Proposition~\ref{prop:size-potential} for branching recursion bound the \emph{number} of activations; rule (9) bounds the cost of a recursive procedure directly, level by level, and the mechanization does not connect the two.

\paragraph{Weights and architectures.}
The logic is parametric in the weights. A cost model for a target is a choice of $\omega$: for a processor on which the duration of an instruction does not depend on the execution history, $\omega$ may be an upper bound of that duration, and the derived bound is then a time bound; for processors with caches or speculative execution no such context-independent weights exist in general (Section~\ref{sec:limitations}), and the loop and recursion bounds derived here are better used as flow facts for a timing analysis of the binary. The companion paper \cite{crosscheckTool} applies the timing schema to the compiled code of a processor of the first kind, with the loop bounds derived on the source; that step relies on the compiler preserving the loops' iteration counts, which is checked there on executions, not proved.

\begin{rem*}[Parallel composition, not formalized]
For fork--join parallelism, the same rules would bound two quantities: the \emph{work}, with the parallel composition of two commands that do not interfere costing the sum of their costs, and the \emph{span}, with parallel composition costing the maximum, as in the analysis of multithreaded computations \cite{blumofe1999scheduling}; the span rule for parallel composition is the rule for a conditional without its test. We do not formalize a parallel composition in the logic; Section~\ref{subsec:concurrency} bounds the duration of a fork--join block of independent tasks on $p$ processors, and of a computation preempted by interrupt handlers, from the cost bounds of the tasks and handlers. Interference between threads through shared state is outside the scope of the framework.
\end{rem*}

\subsection{Certificates and a verified checker}
\label{subsec:checker}

A derivation in the cost logic is a mathematical object. To make it a \emph{certificate} that a program can produce and another program can check, in the manner of proof-carrying code, where the producer of the code supplies a proof with it and the consumer only validates the proof by a simple algorithm \cite{necula1997proof}, we fix a concrete instance in which every side condition is decidable by integer arithmetic.

\paragraph{Linear instance.}
States are valuations of integer variables $x_0, x_1, \dots$. Expressions are affine with integer coefficients; an atomic action is an assignment $x_i := e$, the assignment of an unknown value $x_i := {?}$, or the assignment of an unknown value of which given constraints $g_1 \le 0, \dots$ over the program variables hold afterwards (the value of $y \gg 2$, for instance, is described by $4 x_i \le y \le 4 x_i + 3$); a test is $e \le 0$. An \emph{assertion} is a finite conjunction of constraints $\ell \le 0$, where $\ell$ is affine with integer coefficients in the program variables and in \emph{logical variables} $\gamma_0, \gamma_1, \dots$, which no program modifies; a triple is understood for every value of the logical variables.

\begin{defn}[Linear cost certificate]
\label{def:certificate}
A \emph{symbolic bound} is an expression $c + \sum_j c_j \max(0, \gamma_j) + \sum_{(i,j)} c_{ij} \max(0, \gamma_i) \max(0, \gamma_j)$ with natural numbers $c, c_j, c_{ij}$ over the logical variables, the last sum being a list of \emph{quadratic terms}; it is \emph{linear} if it has no quadratic term, and a constant is a symbolic bound. A \emph{certificate} is a program of the linear instance in which: every sequence carries its intermediate assertion and the (symbolic) bounds of its two components; every conditional carries the bounds of its branches; every loop carries an invariant (an assertion), a potential $e/d$ with $e$ an affine expression and $d \ge 1$ an integer, an upper bound $k + \sum_i k_i \gamma_i$ of the potential on entry ($k \in \mathbb{Z}$, $k_i \in \mathbb{N}$), a logical variable $\gamma_j$ that does not occur in the invariant (and occurs in the bound of the body only in the amortised form below), the bound of its body, and Farkas multipliers for the implication ``invariant and test imply $e \ge d$''; a loop may also have a \emph{conjunctive} test $g_1 \le 0 \wedge \dots \wedge g_m \le 0$, with the same annotations; a loop may be marked \emph{amortised with increasing cost}, with the annotations of a loop with a conjunctive test and a constant entry bound $k$; a loop may instead carry a \emph{logarithmic} potential: an affine expression $e$, an integer base $b \ge 2$, a constant entry bound $k$, a logical variable $\gamma_j$, the (constant) bound of its body, and Farkas multipliers for ``invariant and test imply $e \ge 1$''; or a \emph{growing} logarithmic potential: $e$, $b \ge 2$, a constant $K$, $\gamma_j$, the (constant) bound of its body, and Farkas multipliers for ``invariant and test imply $e \le K$''; and where two further constructs may enclose any part: a \emph{weakening}, which carries an inner precondition, bound and postcondition together with Farkas multipliers for each constraint of the two implications it uses, and a \emph{dead-code} mark, which carries Farkas multipliers showing that the precondition implies $1 \le 0$.
\end{defn}

Farkas multipliers for an implication $\ell_1 \le 0 \wedge \dots \wedge \ell_m \le 0 \Rightarrow \ell \le 0$ are natural numbers $\mu \ge 1$ and $\lambda_1, \dots, \lambda_m$ such that $\mu \ell - \sum_i \lambda_i \ell_i$ is a constant, and this constant is at most $0$; checking them is a comparison of coefficients (soundness direction of Farkas' Lemma, Section~\ref{subsubsec:smt-synthesis}). The \emph{checker} takes a precondition, a certificate, a bound and a postcondition, and proceeds by structural recursion: for an assignment it requires the precondition to be, syntactically, the postcondition with $e$ substituted for $x_i$; for an unknown value, that $x_i$ does not occur in the assertion, and for a constrained one, that $x_i$ does not occur in the precondition and that the postcondition is the precondition followed by the constraints; symbolic bounds are compared coefficient by coefficient, and the quadratic terms of $B_1$ must appear, in the same order, among those of $B_2$ (a syntactic comparison, sound because every term is non-negative), so that $B_1 \le B_2$ holds for every valuation; for a loop with test $g \le 0$, invariant $I$, potential $e/d$ and entry bound $K = k + \sum_i k_i \gamma_i$, that the precondition is $I \wedge e \le K$, that the postcondition is $I \wedge g \ge 1$, that the multipliers prove $e \ge d$ from $I \wedge g \le 0$, that the body is accepted with precondition $I \wedge g \le 0 \wedge e \le \gamma_j$ and postcondition $I \wedge e + d \le \gamma_j$, for a logarithmic loop, that the precondition is $I \wedge e \le k$, that the postcondition is $I$ (with $g \ge 1$ if the test is a single $g \le 0$), that the multipliers prove $e \ge 1$ from $I$ and the test, that the body is accepted with precondition $I \wedge \text{test} \wedge e \le \gamma_j$ and postcondition $I \wedge b\,e \le \gamma_j$, and that $B$ is at least $N(\omega + W) + \omega$ with $N = 0$ if $k \le 0$ and $N = \lfloor \log_b k \rfloor + 1$ otherwise; for a growing loop, the same with precondition $I \wedge e \ge 1$, multipliers proving $e \le K$, body precondition $I \wedge \text{test} \wedge e \ge 1 \wedge e \ge \gamma_j$, body postcondition $I \wedge e \ge b\,\gamma_j$, and $k$ replaced by $K$; and for a linear loop, that the bound $B$ dominates $\lfloor K/d \rfloor \cdot (\omega + W) + \omega$, where $W$ is the bound of the body: either all $k_i$ are $0$ and $B \ge \lfloor k/d \rfloor (\omega + W) + \omega$ coefficient by coefficient (with $\lfloor k/d \rfloor$ read as $0$ when $k < 0$, as in rule (5)), or $W$ is a constant, $B \ge \omega$, and $d (B - \omega)$ dominates $(\omega + W) K$ coefficient by coefficient (in the second form $W$ has no quadratic term); or, third, $d = 1$, $W$ is linear, and $B$ dominates $(\omega + W) \cdot (\max(0, k) + \sum_i k_i \max(0, \gamma_i)) + \omega$, the product of two linear bounds expanded into quadratic terms (a nest of two loops over parameters); in these three forms $\gamma_j$ occurs in no term of $W$; in the third, \emph{amortised} form, all $k_i$ are $0$, the bound of the body is $a + c' \max(0, \gamma_j)$ with no other logical variable, and the constant part of $B$ is at least $F(k)$ of Proposition~\ref{prop:amortised-loop}(1) (compared after doubling, so that no division is needed); for a loop amortised with increasing cost, that the precondition is $I \wedge e \le k$, that the postcondition is as for a logarithmic loop, multipliers proving $e \ge d$, the body accepted with precondition $I \wedge \text{test} \wedge k - e - \gamma_j \le 0 \wedge e + \gamma_j - k \le 0$ (that is, $\gamma_j = k - e$) and postcondition $I \wedge e + \gamma_j - k + d \le 0$, a body bound $a + c' \max(0, \gamma_j)$ with no other logical variable, and a constant part of $B$ of at least $F(k)$ of Proposition~\ref{prop:amortised-loop}(2), doubled; for a loop with a conjunctive test, the same conditions with $g \le 0$ replaced by the conjunction, except that the postcondition is the invariant $I$ alone (the negation of a conjunction is a disjunction, which is not an assertion; an invariant that must hold after the loop, such as a flag recording why the loop ended, is part of $I$); the other cases follow the rules of Definition~\ref{def:derivable-triple}. The logical variable $\gamma_j$ plays the role of the value $v$ in rule (5): the triple about the body holds for every value of $\gamma_j$, in particular for the value of $e$ before the iteration.

\begin{thm}[Soundness of the checker]
\label{thm:checker-soundness}
Assume that every function that the certificate calls meets its specification (Definition~\ref{def:program-certificate}). If the checker accepts a certificate with precondition $P$, bound $B$ and postcondition $Q$, then for every valuation $\rho$ of the logical variables the triple $\{P\}\ c\ \{Q\}\!\downarrow\! B(\rho)$ is valid for the program $c$ of the certificate, and every terminating execution of $c$ from a state satisfying $P$ modifies only the variables that $c$ assigns or that its callees declare. In particular, every execution of $c$ from a state satisfying $P$, terminated or not, costs at most $B(\rho)$.
\end{thm}

\begin{proof}
By induction on the certificate, each case being an instance of the corresponding rule of Theorem~\ref{thm:cost-logic-soundness}, whose side conditions are established as follows. For an assignment $x_i := e$, the value of an affine form $\ell$ after the assignment equals the value of $\ell[e/x_i]$ before it. For an unknown value, the value of $\ell$ does not depend on $x_i$ when the coefficient of $x_i$ in $\ell$ is zero. For the implications, if $\mu \ell = \sum_i \lambda_i \ell_i + r$ with $r \le 0$ a constant, $\lambda_i \ge 0$ and $\mu > 0$, then $\ell_i \le 0$ for all $i$ implies $\mu \ell \le 0$ and hence $\ell \le 0$. For a loop, let the logical variables be given and let $s$ satisfy the invariant and the test. Giving $\gamma_j$ the value $e(s)$ does not affect the invariant, in which $\gamma_j$ does not occur; the precondition of the body holds, so its executions cost at most $W$, and those that terminate do so in a state $t$ satisfying the invariant and $e(t) + d \le e(s)$. Hence $V = e/d$ satisfies the premises of rule (5) with $K(\rho)/d$ as bound of $V$, where $K(\rho) = k + \sum_i k_i \rho_i$, and the loop costs at most $\lfloor K(\rho)/d \rfloor (\omega + W(\rho)) + \omega$; the body's bound $W$ does not depend on $\gamma_j$, which the rule changes. In the first form of the bound condition this is at most $B(\rho)$ coefficient by coefficient; in the second, $\lfloor K(\rho)/d\rfloor \, d \le \max(0, K(\rho))$ and $K(\rho) \le k + \sum_i k_i \max(0, \rho_i)$ since $k_i \ge 0$, so that $d \lfloor K(\rho)/d \rfloor (\omega + W) \le d (B(\rho) - \omega)$; in the third, $d = 1$ and $\max(0, K(\rho)) \le \max(0, k) + \sum_i k_i \max(0, \rho_i)$, and the value of the product of two linear bounds is the product of their values. Executions of the loop from the invariant modify only the variables of the body, by induction on the number of iterations. A loop with a conjunctive test is the same argument for the test ``all $g_i \le 0$'', whose truth implies each constraint $g_i \le 0$; its postcondition $I$ follows from the invariant alone. In the amortised form, giving $\gamma_j$ the value $e(s) \ge d > 0$ shows that an iteration from $s$ costs at most $a + c' e(s)$, and Proposition~\ref{prop:amortised-loop}(1) applies with $V = e$ and $K = k$. For a loop amortised with increasing cost, giving $\gamma_j$ the value $k - e(s)$ satisfies the precondition of the body, so an iteration from $s$ costs at most $a + c' \max(0, k - e(s))$, and a terminating one ends in $t$ with $e(t) + (k - e(s)) - k + d \le 0$, that is $e(t) + d \le e(s)$; Proposition~\ref{prop:amortised-loop}(2) applies with $V = e$ and $K = k$. For a dead-code mark, the precondition is unsatisfiable and rule (7) applies.
For a logarithmic loop, the same argument gives a terminating execution $t$ of the body with $b\,e(t) \le e(s)$ and $e(s) \ge 1$; the function $V(s) = 0$ if $e(s) \le 0$ and $V(s) = \lfloor \log_b e(s) \rfloor + 1$ otherwise then satisfies the premises of rule (5): $V(s) \ge 1$ when the test holds, and $V(t) \le V(s) - 1$, because either $e(t) \le 0$, or $1 \le e(t) \le e(s)/b$, so that $e(s) \ge b$ and $\lfloor \log_b e(t) \rfloor \le \lfloor \log_b e(s) \rfloor - 1$ (the argument of Proposition~\ref{prop:lex-log}(2)); on entry $V \le N$, by monotonicity of the logarithm. For a growing loop, the invariant of rule (5) is $I \wedge e \ge 1$ (after the body, $e(t) \ge b\,e(s) \ge 1$), and $V(s) = N - \lfloor \log_b e(s) \rfloor$: when the test holds, $1 \le e(s) \le K$, so $\lfloor \log_b e(s) \rfloor \le \lfloor \log_b K \rfloor$ and $V(s) \ge 1$; and $e(t) \ge b\,e(s)$ with $e(s) \ge 1$ gives $\lfloor \log_b e(t) \rfloor \ge \lfloor \log_b e(s) \rfloor + 1$, hence $V(t) \le V(s) - 1$.
\end{proof}

\paragraph{Calls, summaries and contracts.}
A certificate for a function states its bound as a function of its parameters: its precondition contains $x_p = \gamma_p$ for each parameter $x_p$, and its symbolic bound is expressed in the $\gamma_p$ (a summary in the sense of Proposition~\ref{prop:summary-call}).

\begin{defn}[Program certificate]
\label{def:program-certificate}
A \emph{program certificate} is a list of functions $F_0, \dots, F_m$, each with a precondition, a symbolic bound, a postcondition, a set of variables that it may modify, and either a certificate (its body) or nothing (an \emph{assumed contract}). A function $F_i$ \emph{meets its specification} in an environment of procedures $\Phi$ if, for every valuation $\rho$ of the logical variables, the triple $\{\mathit{pre}_i\}\ \Phi(i)\ \{\mathit{post}_i\}\!\downarrow\!\mathit{bound}_i(\rho)$ is valid and every terminating execution of $\Phi(i)$ from $\mathit{pre}_i$ modifies only the declared variables of $F_i$. In a body, a \emph{call} of $F_f$ carries program expressions $a_0, a_1, \dots$ for the callee's logical variables (the values of its parameters), Farkas multipliers that prove the callee's precondition with $\gamma_j$ replaced by $a_j$ from the current assertion $P$ (a logical variable of the callee without an expression is replaced by $0$), and, for each $a_j$ whose logical variable occurs in the callee's bound, a symbolic bound $U_j = u + \sum_i u_i \max(0, \gamma_i)$ of the caller with Farkas multipliers for $a_j \le u + \sum_i u_i \gamma_i$, which is at most $U_j$. The checker requires $f$ to be an earlier function, the $U_j$ to be linear, and that the bound dominates $\omega + c^f + \sum_j c^f_j U_j + \sum_{(i,j)} c^f_{ij} U_i U_j$ coefficient by coefficient, where $c^f + \sum_j c^f_j \max(0,\gamma_j) + \sum_{(i,j)} c^f_{ij} \max(0,\gamma_i)\max(0,\gamma_j)$ is the callee's bound and each product $U_i U_j$ is expanded into quadratic terms, and that the postcondition consists of the constraints of $P$ that do not mention the callee's declared variables. A program certificate is accepted if every body is accepted with the functions before it, and modifies only its declared variables.
\end{defn}

\begin{thm}[Soundness of program certificates]
\label{thm:program-soundness}
Let a program certificate be accepted, let $\Phi(i)$ be the program of the body of $F_i$ for every function with a body, and let every assumed contract meet its specification in $\Phi$. Then every function meets its specification in $\Phi$.
\end{thm}

\begin{proof}
By strong induction on $i$. Contracts meet their specifications by hypothesis. For a function with a body, the functions it may call are earlier ones, which meet their specifications by induction; Theorem~\ref{thm:checker-soundness} applies to its body, whose writes are among its declared variables. For the call case of Theorem~\ref{thm:checker-soundness}: in a state $s$ satisfying $P$, let $\rho'_j = a_j(s)$; the callee's precondition holds at $(\rho', s)$ by the Farkas multipliers, so its executions cost at most $c^f + \sum_j c^f_j \max(0, a_j(s)) + \sum_{(i,j)} c^f_{ij} \max(0, a_i(s)) \max(0, a_j(s))$; since $a_j(s) \le U_j(\rho)$ and $U_j(\rho) \ge 0$, $\max(0, a_j(s)) \le U_j(\rho)$, and the products are monotone in non-negative factors; the constraints of $P$ that do not mention the callee's variables still hold after the call, because those variables are the only ones it modifies.
\end{proof}

Recursion is excluded by the order of the functions; rule (9) of the cost logic would allow it, with a rank, but it is not a case of the checker. A contract is trusted: the checker reports it as assumed, and Theorem~\ref{thm:program-soundness} holds relative to it, as Theorem~\ref{thm:cost-logic-soundness} holds relative to the assumed triples $\Gamma$.

The checker is a small program, independent of the analysis that produces certificates: it does not search for invariants, potentials or multipliers, and uses neither an SMT solver nor rational arithmetic. Its soundness is Theorems~\ref{thm:checker-soundness} and~\ref{thm:program-soundness}; an analyzer need not be trusted if its output is checked. Bounds are at most quadratic in the parameters: a loop whose entry bound and body bound both depend on parameters has a certificate when its body bound is linear and its decrease is $1$ (the third form), so a nest of two loops over parameters is covered and a nest of three (a cubic bound) is not; the quadratic terms are compared syntactically, so that a producer must emit the expansion that the checker computes, and the arguments of a call must have linear bounds (a quadratic bound of a callee is instantiated with products of linear ones). With a constant entry bound, a nest in which the entry bound of the inner loop is the potential of the outer one ($\gamma_j$, as for an inner loop from the outer counter to a fixed limit) has the amortised bound; an inner bound that grows while the outer potential decreases (an inner loop from $0$ to the outer counter) has it through the loop amortised with increasing cost, in whose body $\gamma_j$ is what the outer potential has lost. Nests in which the inner bound depends on the outer counter and on a parameter keep the product bound, since the closed forms need a constant $k$. The logarithmic potential of Proposition~\ref{prop:lex-log}(2) is a case of the checker, for loops with constant entry and body bounds, and so is its dual for counters that are multiplied up to a constant limit. The lexicographic potential of Proposition~\ref{prop:lex-log}(1) needs no case of its own when $V_1$ and $V_2$ are affine and $K_2$ is a constant: the combined potential $V_1 (K_2 + 1) + V_2$ is then affine, and the loop has a certificate with the ordinary loop node, whose invariant contains the bounds $V_1 \ge 0$ and $0 \le V_2 \le K_2$. When $K_2$ depends on the parameters the combined potential is a product, which the affine bounds of the checker do not express. Early exits from loops (\texttt{break}, \texttt{continue}, \texttt{return}) need no new construct: a producer can express them with flag variables, a conjunctive test that includes ``flag $\le 0$'', and conditionals that skip the rest of the body once a flag is set; how the plugin does this is described in \cite{crosscheckTool}.

\subsection{Validation against the source program}
\label{subsec:validation}

Theorems~\ref{thm:checker-soundness} and~\ref{thm:program-soundness} bound the cost of the program contained in a certificate. That this program is the function that was analysed is a property of the producer of the certificate, which these theorems do not cover. Following the idea of translation validation \cite{pnueli1998translation,necula2000translation}, the producer is not verified; instead the checker compares the program of a certificate with a second description of the source, produced independently, in a small fragment of C whose meaning is defined in Lean.

\begin{defn}[Mini-C, lowering, validation]
\label{def:minic}
The expressions of \emph{mini-C} are variables $x_i$, integer constants, sums, differences, products, negations and an expression $\top$ that stands for an unknown value; its tests are comparisons $e_1 \bowtie e_2$ with ${\bowtie} \in \{<, \le, >, \ge\}$, negations of tests, and an unknown test; its statements are \texttt{skip}, assignments $x_i := e$, assignments $x_i := e \odot c$ of a quotient, a remainder or a mask by a constant $c$ (${\odot} \in \{/, \bmod, \mathbin{\&}\}$, with $c \ge 1$ for the first two and $c \ge 0$ for a mask; a right shift by $k$ is the quotient by $2^k$), stores $x_i :=_h e$ into an unsigned variable whose largest value is $h$, calls $\mathtt{call}\ f(x_{p_1} := e_1, \dots, x_{p_n} := e_n)$ of the function number $f$ of the program, statements $\mathtt{cost}(\omega)$ that modify no variable, sequences, conditionals, \texttt{while} loops whose test is a conjunction of tests, and \texttt{for} loops. Every assignment, $\mathtt{cost}$ statement, conditional and loop carries a weight $\omega$, the cost of the statement or of the evaluation of its test. The \emph{lowering} $\mathcal{L}$ maps a statement to a program of the checker without annotations (Definition~\ref{def:certificate}):
\begin{itemize}
    \item an expression to its affine form if it has one (a product is affine if one factor is a constant); $e_1 < e_2$ to $e_1 - e_2 + 1 \le 0$, $e_1 \le e_2$ to $e_1 - e_2 \le 0$, and symmetrically for $>$ and $\ge$; the negation of $g \le 0$ to $1 - g \le 0$;
    \item an assignment whose right-hand side has no affine form to $x_i := ?$; $\mathtt{cost}(\omega)$ to an assignment of an unknown value to a variable that no assertion mentions, of weight $\omega$;
    \item $x_i := e \odot c$, for an affine $e$ in which $x_i$ does not occur, to $\mathbf{if}\ (-e \le 0)\ x_i := ?\ \mathrm{with}\ F^{+}\ \mathbf{else}\ x_i := ?\ \mathrm{with}\ F$, a test of weight $0$ followed by an unknown value constrained by facts: $F^{+}$ is $c\,x_i \le e \le c\,x_i + c - 1$ for a quotient, $0 \le x_i \le c - 1$ and $x_i \le e$ for a remainder, $0 \le x_i \le c$ and $x_i \le e$ for a mask; $F$ is empty for a quotient and a remainder (nothing is said for a negative operand) and $0 \le x_i \le c$ for a mask. If $x_i$ occurs in $e$, the statement names a scratch variable $x_t$, different from $x_i$ and from the two dedicated variables below, and its lowering is $x_t := e$ (of weight $0$) followed by the lowering of $x_i := x_t \odot c$ (a scratch variable may be named in any case); if $e$ has no affine form the lowering is $x_i := ?$;
    \item $x_i :=_h e$, for an affine $e$, to $\mathbf{if}\ (e - h \le 0)\ \{\mathbf{if}\ (-e \le 0)\ x_i := e\ \mathbf{else}\ x_i := ?\}\ \mathbf{else}\ x_i := ?$, with tests of weight $0$: the variable receives the value if it is in its range and an unknown value otherwise (and an unknown value if $e$ has no affine form);
    \item a call to the stores of its arguments into the variables $x_{p_j}$ of the callee's parameters, of weight $0$ (stores $:=_h$ for unsigned parameters), followed by the call node of the checker for $f$, of weight $0$, in which the expressions for the callee's logical variables are the variables $x_{p_j}$ (the cost of evaluating the call belongs to the statement that contains it);
    \item a conditional whose test has no affine form to $x_c := ?\,;\ \mathbf{if}\ (x_c \le 0)\ \dots$, for a dedicated variable $x_c$;
    \item a loop with one test to a \texttt{while} loop, with several to a loop with a conjunctive test; $\mathbf{for}\,(i;\ t;\ u)\ b$ to $i\,;\ \mathbf{while}\ t\ (b\,;\ u)$;
\end{itemize}
and $\mathcal{L}$ is undefined on a loop with a conjunct that has no affine form, and on a quotient, remainder or mask whose constant is out of range or whose operand mentions $x_i$ without a scratch variable. The \emph{normal form} $\mathrm{nf}(c)$ of a program $c$ of the checker replaces its annotations (assertions, bounds, invariants, potentials, multipliers) by fixed values (so that a loop with a logarithmic potential becomes the ordinary loop with the same test and body), replaces each consequence or dead-code node by the program it contains, associates sequences to the right, removes \texttt{skip} from sequences, and removes trailing zero coefficients from affine expressions. A certificate with program $c$ is \emph{validated} against a mini-C statement $s$ if $\mathcal{L}(s)$ is defined and $\mathrm{nf}(c) = \mathrm{nf}(\mathcal{L}(s))$.
\end{defn}

The meaning of a mini-C statement $s$ is, by definition, the cost semantics of $\mathcal{L}(s)$ (Definition~\ref{def:valid-triple}): its terminating executions and its execution prefixes, with their costs. Validation is a syntactic test, decided by the checker.

\begin{lem}
\label{lem:nf-equiv}
A program of the checker and its normal form have the same terminating executions and the same execution prefixes, with the same costs.
\end{lem}

\begin{proof}
The translation of a program of the checker into a command of the logic ignores annotations and replaces consequence and dead-code nodes by the program they contain, and a trailing zero coefficient does not change the value of an affine expression. It remains to show that sequential composition is associative and has \texttt{skip} as a unit, for terminating executions and for execution prefixes, and that having the same executions and prefixes is a congruence for sequences, conditionals and loops. Each of these is a case analysis on the derivations of executions and prefixes, by induction on the derivation for loops.
\end{proof}

\begin{thm}[Validated certificates]
\label{thm:validated}
Under the hypotheses of Theorem~\ref{thm:program-soundness}, let the body of the function $F_i$ be validated against the mini-C statement $s$. Then, for every valuation $\rho$ of the logical variables and every state satisfying $\mathit{pre}_i$, every execution prefix of $\mathcal{L}(s)$ costs at most $\mathit{bound}_i(\rho)$, and every terminating execution of $\mathcal{L}(s)$ ends in a state satisfying $\mathit{post}_i$.
\end{thm}

\begin{proof}
By Theorem~\ref{thm:program-soundness}, the body $c$ of $F_i$ satisfies these two properties. Since $\mathrm{nf}(c) = \mathrm{nf}(\mathcal{L}(s))$, Lemma~\ref{lem:nf-equiv}, applied to $c$ and to $\mathcal{L}(s)$, shows that $c$ and $\mathcal{L}(s)$ have the same terminating executions and execution prefixes, with the same costs.
\end{proof}

With validation, the producer of certificates is no longer trusted: what remains is the description of the C function in mini-C, the cost model (the weights, which are part of that description), and the hypotheses under which a C function means its mini-C description. Among the latter: signed integer arithmetic is exact (signed overflow is undefined behaviour in C, assumed absent), variables that the description does not track are read as unknown values and written by $\mathtt{cost}$ statements, and tracked variables are modified only by the assignments that the description shows (their address is not taken). How the companion tool \cite{crosscheckTool} produces the description, and which functions it covers, is described there. Dead code is part of the program: a dead-code node must contain the code of the source that is not reached, which the checker of certificates does not inspect but validation does. Validation is an equality up to the normal form, so a producer that transforms the program, even correctly (for instance by unrolling a loop), is not validated. Among the hypotheses is also that a quotient, remainder or mask of a non-negative value by a constant satisfies the facts $F^{+}$ of Definition~\ref{def:minic}, which is elementary arithmetic, and that the scratch variable of such a statement is not a variable of the source. For a store into an unsigned variable the hypothesis is that the expression is computed modulo $h + 1$ without an intermediate truncation to fewer bits, so that its value over the integers, when it lies in $[0, h]$, is the stored value. A call means the program of the called function in the same certificate, so a validated function is described down to its source only if the functions it calls are validated too; the hypotheses for a call are that the callee modifies no tracked variable of the caller (they are locals whose address is not taken) and that the value of the call is unknown to the caller. A tracked variable may also be a global variable, which the functions of the program share: a call then means the callee's program on the same variables, so its writes are seen, and the hypothesis is that nothing else (an interrupt handler, another thread) writes it while the function runs.

Mini-C has no jumps. They are added as a second fragment, given a meaning of its own, and eliminated by a function that the checker computes; the elimination is proved, not trusted.

\begin{defn}[Mini-C with jumps]
\label{def:minic-jumps}
The statements of \emph{mini-C with jumps} are the statements of mini-C without loops or conditionals (\emph{basic} statements), sequences, conditionals, loops $\mathbf{loop}_\omega(t_1 \wedge \dots \wedge t_m)\ \{b\}\ u$ with a body $b$ and an optional increment $u$ (a basic statement; a \texttt{for} loop is its initialisation followed by such a loop), $\mathtt{break}_\omega$, $\mathtt{continue}_\omega$, $\mathtt{return}\ s$ for a basic statement $s$ (the evaluation of the returned value), $\mathtt{switch}_\omega(g_1, \dots, g_k)$ with case groups $g_1, \dots, g_k$, and $\mathbf{do}\ \{b\}\ \mathbf{while}_\omega(t_1 \wedge \dots \wedge t_m)$. Its meaning is a big-step semantics with \emph{outcomes}: a statement started in a state ends in a state with the outcome \emph{normal}, \emph{break}, \emph{continue} or \emph{return}, or is \emph{interrupted} (its cost so far, which gives the execution prefixes). A basic statement means its lowering; a sequence stops at the first outcome that is not normal; a loop whose test holds runs its body, then, after a normal end or a \texttt{continue}, its increment and the loop again, and it ends normally after a \texttt{break} of its body and with \emph{return} after a \texttt{return}; a \texttt{switch} pays $\omega$, then runs the groups $g_i, g_{i+1}, \dots$ in sequence for some $i$ (or none), up to a \texttt{break}, which ends the \texttt{switch} normally; a \texttt{do} loop runs its body, then, after a normal end or a \texttt{continue}, pays $\omega$ for its test and goes on while it holds. Which group is entered is not modelled (the value tested is not compared with the labels), so every execution of the C statement is one of the \texttt{switch}.
\end{defn}

The \emph{elimination} $\mathcal{E}$ maps a statement with jumps to a mini-C statement, by the usual flag variables: a loop with a \texttt{break} or a \texttt{return} in its body sets their flags to $0$, its test requires them to be $0$ as well, its body first clears the flag of \texttt{continue}, and its increment runs only if the flags of \texttt{break} and \texttt{return} are $0$; a jump stores $1$ into its flag; what follows a statement that may jump is moved into the branches that do not jump, and what follows a loop or a \texttt{switch} that may set the flag of \texttt{return}, or of a \texttt{continue} of an enclosing loop, runs only if that flag is $0$ (tests of weight $0$); a \texttt{switch} is its cost followed by each group $g_j$ under $x_s := ?\,;\ \mathbf{if}\ (x_s \le 0)\ g_j$. A \texttt{do} loop whose body has no \texttt{break} or \texttt{continue} of its own is eliminated as its body followed by the loop $\mathbf{while}(t_1 \wedge \dots \wedge t_m)$ with the same body. $\mathcal{E}$ is computed by the checker with a continuation (the statements that follow, already eliminated) and a stack of the enclosing loops and \texttt{switch} statements; it is undefined if a flag or the variable $x_s$ of a \texttt{switch} (\emph{ghost variables}) is named by a statement of the source, is not fresh along the nesting of loops and \texttt{switch} statements, or is one of the two dedicated variables, if a loop left by a jump has no test of its own, or if a basic statement calls a function that is not earlier in the program. A certificate with program $c$ is validated against a statement $j$ with jumps if $c$ is validated against $\mathcal{E}(j)$.

\begin{thm}[Validated certificates with jumps]
\label{thm:validated-jumps}
Under the hypotheses of Theorem~\ref{thm:program-soundness}, let the body of the function $F_i$ be validated against the statement with jumps $j$, and let no function before $F_i$ read or write a ghost variable of $j$. Then, for every valuation $\rho$ of the logical variables and every state $\sigma$ satisfying $\mathit{pre}_i$, every interrupted execution of $j$ from $\sigma$, in the semantics of Definition~\ref{def:minic-jumps}, costs at most $\mathit{bound}_i(\rho)$, and every execution of $j$ from $\sigma$ that ends normally or by \texttt{return} ends in a state that agrees, outside the ghost variables and the variable written by $\mathtt{cost}$ statements, with a state satisfying $\mathit{post}_i$.
\end{thm}

\begin{proof}
By induction on the derivation of the execution of $j$, generalised over the context (the enclosing loops and \texttt{switch} statements) and the continuation $k$, we show: from every state that agrees with the initial one outside the ghost variables and the cost variable and in which the flags of the context are $0$, the lowering of $\mathcal{E}(j)$ followed by $k$ has an execution that costs at least as much and ends in a state that again agrees outside those variables; after a normal end the flags of the context are $0$ and the run goes on with $k$; after a jump the flag of that jump is $1$, the other flags of the context are $0$, and $k$ is not run; an interrupted execution is matched by an execution prefix. Basic statements and tests do not name the ghost variables, so they run in the same way from states that agree outside them; so do the called functions, which are earlier functions by hypothesis. For a loop a second statement, about the loop command itself after its flags are cleared, is proved in the same induction; after a \texttt{break} or a \texttt{return} the lowered loop evaluates its test once more than the loop with jumps, which only adds cost. For a \texttt{switch} the groups before the one entered are skipped by choosing $x_s = 1$, at no cost, and after a jump the following groups are skipped in the same way. A statement with \texttt{do} loops is first desugared ($\mathbf{do}\ \{b\}\ \mathbf{while}(t)$ into $b\,;\ \mathbf{while}(t)\ b$): every execution of it is an execution of the desugared statement when the bodies of its \texttt{do} loops have no \texttt{break} or \texttt{continue} of their own, by induction on the derivation, and $\mathcal{E}$ gives the same result on both. Theorem~\ref{thm:validated}, applied to $\mathcal{E}(j)$ with the context empty and $k = \mathtt{skip}$, then gives the bound and the postcondition.
\end{proof}

Extending mini-C with a statement whose lowering uses the existing programs of the checker means extending $\mathcal{L}$ only (Lemma~\ref{lem:nf-equiv} and Theorem~\ref{thm:validated} are about the programs of the checker), while a new kind of program would also need a case of the lemma.

\subsection{Interrupts and fork-join}
\label{subsec:concurrency}

The bounds of the previous sections are costs of sequential code. Real-time software combines such code in two ways that this subsection addresses: a computation is preempted by interrupt handlers, and independent tasks are run in parallel. In both cases the duration is bounded from the cost bounds of the components, under the hypothesis that one unit of cost takes one unit of time (a cost model that is a timing model, Section~\ref{subsec:cost-logic}). Both results are classical; what is new here is only their connection to the certified bounds, and their mechanization in our setting (a mechanized theory of response-time analysis, far more general, is Prosa \cite{cerqueira2016prosa}).

\begin{defn}[Computation under interrupts]
\label{def:interrupt-model}
Time is divided into slots $0, 1, 2, \dots$ from the start of a computation that needs $C$ slots. There are handlers $k \in K$; each activation of handler $k$ needs at most $c_k$ slots, activations of $k$ are released at least $T_k > 0$ slots apart (there may be infinitely many), and none is released before slot $0$. A \emph{schedule} assigns to each slot either nothing, the computation, or an activation of a handler released at or before that slot, with at most $c_k$ slots per activation; it is \emph{admissible} if no slot is idle while the computation has received fewer than $C$ slots.
\end{defn}

\begin{lem}
\label{lem:sporadic}
At most $\lceil t / T_k \rceil$ activations of handler $k$ are released before slot $t$, and handler $k$ receives at most $c_k \lceil t / T_k \rceil$ of the first $t$ slots.
\end{lem}

\begin{proof}
Two activations released before $t$ are at least $T_k$ apart, so they lie in different intervals $[jT_k, (j+1)T_k)$, of which $\lceil t/T_k \rceil$ meet $[0, t)$. Each slot of handler $k$ belongs to one of these activations, and each activation receives at most $c_k$ slots.
\end{proof}

\begin{thm}[Response time]
\label{thm:response-time}
In every admissible schedule, the computation receives its $C$ slots within the first $R$ slots, for every $R$ such that
\[ C + \sum_{k \in K} c_k \left\lceil \frac{R}{T_k} \right\rceil \le R. \]
\end{thm}

\begin{proof}
Suppose the computation receives fewer than $C$ of the first $R$ slots. Then none of these slots is idle, so $R$ is at most the number of slots of the computation, which is less than $C$, plus the number of slots of the handlers, which is at most $\sum_k c_k \lceil R/T_k \rceil$ by Lemma~\ref{lem:sporadic}. This contradicts the hypothesis.
\end{proof}

This is the response-time bound of fixed-priority preemptive scheduling \cite{joseph1986finding,audsley1993applying}, for one computation of the lowest priority; the least such $R$ is found by the iteration $R_0 = C$, $R_{n+1} = C + \sum_k c_k \lceil R_n / T_k \rceil$, stopped at a fixed point, which is Joseph and Pandya's function \emph{Response} \cite{joseph1986finding}. Their analysis starts at the critical instant of Liu and Layland, when inputs of every level arrive together; Theorem~\ref{thm:response-time} bounds instead every schedule in which no activation is pending at slot~$0$. The priority order of the handlers is not needed for this bound, only that the processor does not idle while the computation is pending. The hypothesis that no activation is pending at slot $0$ is the restrictive one: the classical analysis removes it by considering the busy period that contains the start of the computation, an argument that we do not state or mechanize here. With $C$ and the $c_k$ certified bounds of the computation and of the handlers (Theorem~\ref{thm:program-soundness}), plus a bound on the cost of entering and leaving a handler, the only assumptions left are the minimum inter-arrival times $T_k$, the overhead, and the hypothesis that cost is time. The checker accepts a response-time claim when the task and handlers have accepted certificates with constant bounds and the inequality of Theorem~\ref{thm:response-time} holds.

\begin{prop}[Fork-join, after Graham]
\label{prop:graham}
Let independent tasks of costs $w_1, \dots, w_n$ be scheduled on $p \ge 1$ processors by a greedy scheduler, which starts each task, in a fixed order, on a processor of least load. Then every processor's load $L$ satisfies $p L \le \sum_i w_i + p \max_i w_i$, that is, the makespan is at most $W/p + S$, where $W = \sum_i w_i$ is the work and $S = \max_i w_i$ the span; $W/p$ and $S$ are lower bounds on the makespan of every schedule. If $w_i \le B_i$ for every $i$, then $p L \le \sum_i B_i + p \max_i B_i$.
\end{prop}

\begin{proof}
By induction on the number of tasks scheduled, the loads sum to the work of the tasks scheduled so far and satisfy the inequality. When task $w$ is started on a processor of least load $\ell$, $p\ell$ is at most the sum of the loads, which is the work so far; the new load $\ell + w$ thus satisfies $p(\ell + w) \le W' + pw$, where $W'$ is the new work, and the other loads are unchanged while the right-hand side does not decrease. The last statement follows from the monotonicity of sums and maxima.
\end{proof}

Proposition~\ref{prop:graham} is Graham's bound for list scheduling \cite{graham1966bounds}, here for independent tasks; with dependencies between tasks the same form $W/p + S$ holds for greedy schedulers of task graphs, where $S$ is the longest path (the greedy-scheduling theorem, stated in \cite{blumofe1999scheduling} as an adaptation of results of Brent and of Graham), which we do not mechanize. With certified bounds $B_i$ of the tasks, the duration of a fork--join block of independent tasks is at most $\sum_i B_i / p + \max_i B_i$, under the hypotheses that the tasks do not interfere (no shared variables, locks or shared caches), that cost is time, and that the costs of forking and joining are added.

\subsection{Stack depth}
\label{subsec:stack}

Cost adds up along an execution. The stack that a program needs does not: it is a high-water mark, the largest depth reached, and the rule for composing it over calls is a maximum where the rule for cost is a sum.

\begin{defn}[Activation trees and their stack]
\label{def:activation-tree}
An \emph{activation tree} is a tree whose nodes are labelled by functions and whose edges are labelled by natural numbers: a node $f$ with an edge $d$ to a subtree $t$ is an activation of $f$ that calls, when its stack has depth $d$, the function at the root of $t$. Given the largest depth $\mathrm{own}(f)$ that the code of each function $f$ reaches by itself, the \emph{stack} of an activation tree is
\[
  \mathrm{hw}(f;\ (d_1, t_1), \dots, (d_n, t_n)) \;=\; \max\bigl(\mathrm{own}(f),\ \max_i\, (d_i + \mathrm{hw}(t_i))\bigr).
\]
\end{defn}

\begin{prop}[Stack bounds]
\label{prop:stack}
\begin{enumerate}
    \item Let a program be given by its call sites: triples $(f, d, g)$, meaning that $f$ may call $g$ at depth $d$. If $B$ maps functions to natural numbers with $\mathrm{own}(f) \le B(f)$ for every $f$ and $d + B(g) \le B(f)$ for every call site $(f, d, g)$, then every activation tree all of whose edges are call sites has stack at most $B$ of its root.
    \item If $\mathrm{own}(f) \le F$ for every $f$ and every call is made at depth at most $F$, an activation tree of height $n$ has stack at most $(n + 1) F$.
\end{enumerate}
\end{prop}

\begin{proof}
Both by induction on the tree. (1) For a node $f$, each term $d_i + \mathrm{hw}(t_i)$ is at most $d_i + B(g_i) \le B(f)$ by the induction hypothesis and the condition on the call site, and $\mathrm{own}(f) \le B(f)$. (2) A subtree $t_i$ has height at most $n - 1$, so $d_i + \mathrm{hw}(t_i) \le F + n F$.
\end{proof}

For a program without recursion the least $B$ of the first statement is computed bottom-up on the call graph; the second statement turns a bound on the depth of a recursion (Proposition~\ref{prop:tree-envelope}) into a stack bound. The companion paper applies the first to compiled code, where $\mathrm{own}$ and the depths of call sites are read from the control-flow graph of the binary; that the executions of a machine program are described by such trees is a property of the processor model, not proved here.

\section{Examples}
\label{sec:examples}

The following examples are simple enough for the equation systems to be solved in closed form, so that the results can be checked by hand. The method is that of an earlier prototype implementation of the rules of Section~\ref{sec:application} (2008--2017, directory \lean{history/revenge} of the repository), which analyzes the source code without executing it and whose tests include the infinite loop and the linear recursion below; the systems are given with minor simplifications for readability. Solving each system is equivalent to executing the code, and each solution has additionally been verified in Lean~4 against the big-step semantics (Section~\ref{sec:mechanization}).

\subsection{Parametric while loop}

This example shows a while loop whose behavior depends on a parametric (unknown) initial value $R$.

\begin{lstlisting}
int i = random();
int x = 1; int y = 2;
while (i++ < 13) {
    i = i + 1;
    x = y + 2;
    y = x + 3;
}
\end{lstlisting}

The guard has a side effect, and the variables $x, y$ do not influence the guard but modify each other at every cycle. One full iteration (successful guard and body) maps $(i, x, y)$ to $(i + 2, y + 2, y + 5)$, and the failing guard increments $i$ once more. Writing $n \ge 1$ for the number of iterations, $i_{loop}$ for the value compared by the last successful guard and $i_{last}^{0}$ for the value compared by the failing guard, the generated system is:

\[
\begin{cases}
i_{init} & =R\\
y_{loop}^{0} & =2\\
x_{loop}^{0} & =y_{loop}^{0}+2\\
i_{init} & <13\\
i_{loop} & =i_{init}+2(n-1)\\
i_{loop} & <13\\
i_{last}^{0} & =i_{loop}+2\\
i_{last}^{0} & \geq13\\
i_{last}^{1} & =i_{last}^{0}+1\\
x_{loop}^{1} & =y_{loop}^{0}+5n-3\\
y_{loop}^{1} & =x_{loop}^{1}+3
\end{cases}
\]

The two inequalities $R + 2(n-1) < 13 \le R + 2n$ have exactly one integer solution $n$, which is therefore the minimum of Proposition~\ref{prop:min_x}. If the loop is entered at least once ($R<13$):

\[
\begin{cases}
n & =\left\lceil \frac{13-R}{2}\right\rceil \\
i & =i_{last}^{1}=R+2n+1\\
x & =x_{loop}^{1}=5n-1\\
y & =y_{loop}^{1}=5n+2
\end{cases}
\]

For instance $R = 0$ gives $n = 7$ and $(i, x, y) = (15, 34, 37)$. If the loop is never entered ($R\geq13$), only the side effect of the failing guard remains:

\[
\begin{cases}
i & =R+1\\
x & =1\\
y & =2
\end{cases}
\]

\begin{rem*}
The value compared by the failing guard is $i_{last}^{0}=R+2n$ if $R<13$, and $R$ otherwise.
\end{rem*}

\paragraph{Relaxed analysis.}
The function $V(i, x, y) = 13 - i$ is a ranking potential with $\Delta = 2$: on continuing states $i < 13$, so $V > 0$, and each iteration decreases $V$ by exactly $2$. Theorem~\ref{thm:lyapunov-bound} gives $\bar{n} \le \max(0, \lceil (13 - R)/2 \rceil)$, which coincides with the exact iteration count: the envelope is tight on this example. With unit weights and four instructions per iteration (guard and three assignments), the cost of the loop is at most $4 \cdot \lceil (13 - R)/2 \rceil$ plus one final guard evaluation.

\subsection{Infinite while loop}

This example shows a never-ending loop:

\begin{lstlisting}
int i = 0;
while (i != 1) {
    i = 0;
}
\end{lstlisting}

The generated equations are:

\[
\begin{cases}
i_{init} & =0\\
i_{init} & \neq1\\
i_{loop} & =0\\
i_{loop} & \neq1\\
i_{last} & =i_{loop}\\
i_{last} & =1
\end{cases}
\]

The system is infeasible, so the loop is classified as non-terminating: no iterate of the body exits, and by Theorem~\ref{thm:extensional-equivalence} the loop has no big-step derivation. Consistently, no ranking potential exists, since the body maps the continuing state $i = 0$ to itself.

\subsection{Linear recursion}

\begin{lstlisting}
int foo(int a) {
    if (a > 4)
        a = foo(a - 2) + 3;
    return a + 6;
}
foo(10);
\end{lstlisting}

In the form of Proposition~\ref{prop:linrec}, $\text{fwd}(a) = a - 2$, $\text{stop}(a) = a + 6$ and $\text{back}(r) = r + 3 + 6$ (the assignment followed by the fall-through return); the recursion is frame-independent because the caller's $a$ is overwritten. The generated equations, after minor simplifications, are:

\[
\begin{cases}
a_{init} & =10\\
a_{init} & >4\\
a_{loop}^{1} & =a_{init}-2n\\
a_{last} & =a_{loop}^{1}\\
a_{last} & \leq4\\
a_{stop} & =a_{last}+6\\
a_{loop}^{2} & =a_{stop}+3n\\
a_{back} & =a_{loop}^{2}+6n
\end{cases}
\]

Since the input is known, the numerical values can be computed. The depth is the least $n$ with $10 - 2n \le 4$:

\[
\begin{cases}
n & =3\\
a & =a_{back}=37
\end{cases}
\]

\begin{rem*}
The descending part of the recursion ends at $a_{last}=4$.
\end{rem*}

\subsection{Non-linear recursion}
This example is a symmetric binary recursive countdown:

\begin{lstlisting}
int foo(int x) {
  if (x > 0)
    return foo(x - 1) + foo(x - 1);
  return 1;
}
foo(2);
\end{lstlisting}

The analysis generates a system of equations indexed by tree positions $\pi \in \{\varepsilon, 0, 1, 00, 01, 10, 11\}$:

\[
\begin{cases}
x_{\varepsilon} = 2, & x_{\varepsilon} > 0 \\
x_{0} = x_{\varepsilon} - 1 = 1, & x_{1} = x_{\varepsilon} - 1 = 1 \\
x_{0} > 0, & x_{1} > 0 \\
x_{00} = x_{0} - 1 = 0, & x_{01} = x_{0} - 1 = 0 \\
x_{10} = x_{1} - 1 = 0, & x_{11} = x_{1} - 1 = 0 \\
x_{00} \le 0 \implies v_{00} = 1, & x_{01} \le 0 \implies v_{01} = 1 \\
x_{10} \le 0 \implies v_{10} = 1, & x_{11} \le 0 \implies v_{11} = 1 \\
v_{0} = v_{00} + v_{01} = 2, & v_{1} = v_{10} + v_{11} = 2 \\
v_{\varepsilon} = v_{0} + v_{1} = 4
\end{cases}
\]

The solution is the minimal computation tree $\bar{T} = B^{\otimes 2}$, the full binary tree of depth $d = 2$. The total work is
\[
\cost(\bar{T}) = 2^{d+1} - 1 = 7 \text{ calls},
\]
and the value synthesized at the root is $v_{\varepsilon} = 4$. The function $V_{\text{tree}}(x) = \max(x, 0)$ is a depth ranking potential, so Proposition~\ref{prop:tree-envelope} bounds the work of $\text{foo}(x)$ by $2^{x+1} - 1$ for every $x \ge 0$; here the envelope is attained.

\section{Mechanization in Lean 4}
\label{sec:mechanization}

The development accompanying this paper consists of about 11{,}000 lines of Lean~4 \cite{demoura2021lean4} over mathlib \cite{mathlib2020}, with 499 theorems; the modules of the third paper \cite{crosscheckBinary} (about 5{,}000 lines, 193 theorems) are in the same repository. It contains no \lean{sorry} and declares no axioms; the Lean command \lean{\#print axioms} reports, for every theorem cited here, at most the three standard axioms of Lean's logic (\lean{propext}, \lean{Quot.\allowbreak sound}, \lean{Classical.\allowbreak choice}). Table~\ref{tab:lean} lists the correspondence between the statements of the paper and the Lean theorems.

\begin{table}[p]
\centering
\scriptsize
\setlength{\tabcolsep}{3pt}
\begin{tabular}{@{}>{\raggedright\arraybackslash}p{0.205\textwidth}>{\raggedright\arraybackslash}p{0.61\textwidth}>{\raggedright\arraybackslash}p{0.15\textwidth}@{}}
\toprule
Paper statement & Lean theorem(s) & Module \\
\midrule
Eq.~\eqref{eq:iter-hom} (valuation homomorphism) & \lean{evalTrace\_\allowbreak concat}, \lean{evalTrace\_\allowbreak iter} & \lean{FixedPoint} \\
Def.~\ref{def:subadditive-norm} (weighted cost) & \lean{wcost\_\allowbreak append}, \lean{wcost\_\allowbreak iter}, \lean{wcost\_\allowbreak one} & \lean{Trace} \\
Prop.~\ref{prop:poset-properties} & \lean{order\_\allowbreak refl/\allowbreak trans/\allowbreak antisymm}, \lean{order\_\allowbreak cost\_\allowbreak le} & \lean{Trace} \\
Prop.~\ref{prop:n-monotony} & \lean{iter\_\allowbreak monotone\_\allowbreak iff} & \lean{Trace} \\
Thm.~\ref{thm:minimal-trace} & \lean{exists\_\allowbreak minimal\_\allowbreak iteration} & \lean{Trace} \\
Prop.~\ref{prop:worst-case-step} & \lean{runTrace\_\allowbreak cost\_\allowbreak le}, \lean{runTrace\_\allowbreak wcost\_\allowbreak le} & \lean{WhileLoop} \\
Prop.~\ref{prop:nested-cost} & \lean{nested\_\allowbreak trace\_\allowbreak cost}, \lean{sequential\_\allowbreak loops\_\allowbreak cost/\allowbreak prefix} & \lean{Trace} \\
Prop.~\ref{prop:tropical-dominance}, \ref{prop:tropical-collapse} & \lean{condTrace\_\allowbreak cost\_\allowbreak le}, \lean{branch\_\allowbreak supremum\_\allowbreak bound} & \lean{Dominance} \\
Def.~\ref{def:subadditive-norm}, Props.~\ref{prop:worst-case-step}, \ref{prop:tropical-dominance}, \ref{prop:tropical-collapse}, Thm.~\ref{thm:lyapunov-bound}, for every subadditive norm & \lean{CostNorm}, \lean{CostNorm.\allowbreak runTrace\_\allowbreak le}, \lean{CostNorm.\allowbreak condTrace\_\allowbreak le}, \lean{CostNorm.\allowbreak blocks\_\allowbreak le}, \lean{CostNorm.\allowbreak wcet\_\allowbreak bound} & \lean{Norms} \\
Prop.~\ref{prop:poset-properties}, Thm.~\ref{thm:minimal-trace}, Prop.~\ref{prop:min_x} ($\epsilon$, well-foundedness, least trace) & \lean{order\_\allowbreak nil}, \lean{order\_\allowbreak wf}, \lean{minimal\_\allowbreak iteration\_\allowbreak least}, \lean{runTrace\_\allowbreak prefix} & \lean{Norms} \\
Thm.~\ref{thm:extensional-equivalence} & \lean{while\_\allowbreak trace\_\allowbreak semantics} & \lean{WhileLoop} \\
Prop.~\ref{prop:min_x} & \lean{whileBigStep\_\allowbreak iff\_\allowbreak minimalExit}, \lean{MinimalExit.\allowbreak unique} & \lean{WhileLoop} \\
Prop.~\ref{prop:tree-order-properties} & \lean{treePrefix\_\allowbreak refl/\allowbreak trans/\allowbreak antisymm}, \lean{treePrefix\_\allowbreak size\_\allowbreak le}, \lean{treePrefix\_\allowbreak depth\_\allowbreak le}, \lean{treePrefix\_\allowbreak graft}, \lean{treePrefix\_\allowbreak fullTree}, \lean{size\_\allowbreak le\_\allowbreak of\_\allowbreak depth\_\allowbreak le} & \lean{Tree} \\
Prop.~\ref{prop:tree-power-complexity} & \lean{treePow\_\allowbreak leafCount}, \lean{treePow\_\allowbreak size}, \lean{treePow\_\allowbreak cherry}, \lean{treePow\_\allowbreak chain}, \lean{graft\_\allowbreak size}, \lean{graft\_\allowbreak assoc} & \lean{Tree} \\
Prop.~\ref{prop:linrec} & \lean{framedBigStep\_\allowbreak iff}, \lean{linRec\_\allowbreak iff}, \lean{linear\_\allowbreak recursion\_\allowbreak cost} & \lean{Recursion} \\
Prop.~\ref{prop:tree-optimization} & \lean{exec\_\allowbreak least}, \lean{exec\_\allowbreak unique}, \lean{exec\_\allowbreak size\_\allowbreak min}, \lean{treeBigStep\_\allowbreak iff} & \lean{Recursion} \\
Thm.~\ref{thm:lyapunov-bound} & \lean{ranking\_\allowbreak bound}, \lean{ranking\_\allowbreak terminates}, \lean{wcet\_\allowbreak bound} & \lean{Ranking} \\
Prop.~\ref{prop:tree-envelope} & \lean{exec\_\allowbreak depth\_\allowbreak le}, \lean{exec\_\allowbreak size\_\allowbreak le} & \lean{Recursion} \\
Thm.~\ref{thm:lyapunov-bound} (relational) & \lean{ranking\_\allowbreak bound\_\allowbreak floor}, \lean{ranking\_\allowbreak bound\_\allowbreak of\_\allowbreak le}, \lean{partial\_\allowbreak ranking\_\allowbreak floor} & \lean{Certificate} \\
\bottomrule
\end{tabular}
\caption{Correspondence between the paper and the Lean~4 development (continued below).}
\label{tab:lean}
\end{table}

\begin{table}[p]
\centering
\scriptsize
\setlength{\tabcolsep}{3pt}
\begin{tabular}{@{}>{\raggedright\arraybackslash}p{0.205\textwidth}>{\raggedright\arraybackslash}p{0.61\textwidth}>{\raggedright\arraybackslash}p{0.15\textwidth}@{}}
\toprule
Paper statement & Lean theorem(s) & Module \\
\midrule
Def.~\ref{def:tree-algebra}, \ref{def:tree-order}, \ref{def:tree-exponentiation}, Props.~\ref{prop:tree-order-properties}, \ref{prop:tree-power-complexity}, \ref{prop:tree-optimization} for every branching factor $b$ & \lean{KTree}, \lean{KTree.\allowbreak le\_\allowbreak refl/\allowbreak trans/\allowbreak antisymm}, \lean{size\_\allowbreak mono}, \lean{depth\_\allowbreak mono}, \lean{le\_\allowbreak graft}, \lean{le\_\allowbreak full}, \lean{size\_\allowbreak full}, \lean{size\_\allowbreak le\_\allowbreak closed}, \lean{graft\_\allowbreak leaves}, \lean{graft\_\allowbreak size}, \lean{pow\_\allowbreak leaves}, \lean{pow\_\allowbreak size\_\allowbreak closed}, \lean{pow\_\allowbreak size\_\allowbreak rat}, \lean{pow\_\allowbreak step}, \lean{pow\_\allowbreak chain}; \lean{KRecurrence.\allowbreak exec\_\allowbreak least}, \lean{exec\_\allowbreak unique}, \lean{exec\_\allowbreak size\_\allowbreak min}, \lean{exec\_\allowbreak depth\_\allowbreak min}, \lean{bigStep\_\allowbreak iff}; example \lean{tern\_\allowbreak exec} & \lean{Branching} \\
Def.~\ref{def:valid-triple}, \ref{def:derivable-triple} & \lean{Cmd}, \lean{Exec}, \lean{Pre}, \lean{Valid}, \lean{Specs}, \lean{Derivable} & \lean{Language} \\
Thm.~\ref{thm:cost-logic-soundness} & \lean{soundness}, \lean{valid\_\allowbreak seq}, \lean{valid\_\allowbreak ite}, \lean{valid\_\allowbreak while}, \lean{valid\_\allowbreak call}, \lean{valid\_\allowbreak conseq}, \lean{derivable\_\allowbreak cost\_\allowbreak le}; examples \lean{countLoop\_\allowbreak cost}, \lean{down\_\allowbreak cost} (recursion) & \lean{Language} \\
Prop.~\ref{prop:no-divergence} & \lean{Div}, \lean{PosWeights}, \lean{div\_\allowbreak prefixes\_\allowbreak unbounded}, \lean{valid\_\allowbreak no\_\allowbreak div}, \lean{derivable\_\allowbreak no\_\allowbreak div}; example \lean{spin\_\allowbreak no\_\allowbreak bound} & \lean{Divergence} \\
Def.~\ref{def:certificate} & \lean{ACmd}, \lean{Lin}, \lean{PExp}, \lean{check}, \lean{entails1} & \lean{Checker} \\
Thm.~\ref{thm:checker-soundness} & \lean{check\_\allowbreak sound}, \lean{loop\_\allowbreak sound}, \lean{logloop\_\allowbreak sound}, \lean{growloop\_\allowbreak sound}, \lean{loopBoundOk\_\allowbreak sound}, \lean{triloop\_\allowbreak sound}, \lean{uploop\_\allowbreak sound}, \lean{certified\_\allowbreak cost\_\allowbreak le}, \lean{entails1\_\allowbreak sound}, \lean{while\_\allowbreak frame}; examples \lean{growCert\_\allowbreak cost}, \lean{triCert\_\allowbreak cost}, \lean{upCert\_\allowbreak cost}, \lean{prodCert\_\allowbreak cost} & \lean{Checker} \\
Def.~\ref{def:program-certificate}, Thm.~\ref{thm:program-soundness} & \lean{FunCert}, \lean{FunSpec}, \lean{checkProgram}, \lean{program\_\allowbreak sound}; example \lean{exProgram\_\allowbreak checks}, \lean{exProgram\_\allowbreak 17\_\allowbreak rejected}, \lean{exMain\_\allowbreak cost} & \lean{Checker} \\
Def.~\ref{def:minic}, Lem.~\ref{lem:nf-equiv}, Thm.~\ref{thm:validated} & \lean{lower}, \lean{nf}, \lean{validate}, \lean{nf\_\allowbreak equiv}, \lean{validated\_\allowbreak sound}; example \lean{countCert\_\allowbreak validated} & \lean{MiniC} \\
Def.~\ref{def:minic-jumps}, Thm.~\ref{thm:validated-jumps} & \lean{JExecD}, \lean{elim}, \lean{validateJ}, \lean{sim}, \lean{desugar\_\allowbreak exec}, \lean{validatedJ\_\allowbreak sound} & \lean{MiniCJump} \\
Def.~\ref{def:interrupt-model}, Lem.~\ref{lem:sporadic}, Thm.~\ref{thm:response-time} & \lean{Admissible}, \lean{sporadic\_\allowbreak count}, \lean{handler\_\allowbreak demand}, \lean{response\_\allowbreak time\_\allowbreak bound}, \lean{rtaHolds\_\allowbreak sound}; examples \lean{oneRun\_\allowbreak completes}, \lean{perRun\_\allowbreak completes} (a handler activated forever) & \lean{Interrupts} \\
Prop.~\ref{prop:graham} & \lean{Greedy}, \lean{greedy\_\allowbreak bound}, \lean{greedy\_\allowbreak exists}, \lean{forkjoin\_\allowbreak bound}, \lean{makespan\_\allowbreak lower} & \lean{ForkJoin} \\
Prop.~\ref{prop:lex-log} & \lean{lex\_\allowbreak potential}, \lean{lex\_\allowbreak ranking\_\allowbreak bound}, \lean{log\_\allowbreak ranking\_\allowbreak bound} & \lean{Certificate} \\
Prop.~\ref{prop:stack} & \lean{highWater\_\allowbreak le\_\allowbreak of\_\allowbreak bound}, \lean{highWater\_\allowbreak le\_\allowbreak depth} & \lean{Stack} \\
Prop.~\ref{prop:summary-call} & \lean{validSym\_\allowbreak iff}, \lean{validSym\_\allowbreak call}, \lean{validSym\_\allowbreak seq} & \lean{Language} \\
Prop.~\ref{prop:context-bound} & \lean{run\_\allowbreak invariant}, \lean{ranking\_\allowbreak bound\_\allowbreak floor\_\allowbreak of\_\allowbreak invariant}, \lean{ranking\_\allowbreak bound\_\allowbreak of\_\allowbreak context}, \lean{call\_\allowbreak bound\_\allowbreak of\_\allowbreak context} & \lean{Certificate} \\
Sec.~\ref{subsubsec:smt-synthesis} (Farkas) & \lean{farkas\_\allowbreak sound}, \lean{farkas\_\allowbreak sound\_\allowbreak affine} & \lean{Certificate} \\
Loop cost formula; Prop.~\ref{prop:amortised-loop} & \lean{while\_\allowbreak cost\_\allowbreak bound}; \lean{valid\_\allowbreak whileSum}, \lean{valid\_\allowbreak whileMono} & \lean{Certificate}, \lean{Language} \\
Prop.~\ref{prop:size-potential} & \lean{callTree\_\allowbreak size\_\allowbreak le\_\allowbreak potential}, \lean{callTree\_\allowbreak size\_\allowbreak le\_\allowbreak geom}, \lean{geom\_\allowbreak closed} & \lean{Arity} \\
Section~\ref{sec:examples} & \lean{parametric\_\allowbreak loop\_\allowbreak minimal}, \lean{parametric\_\allowbreak loop\_\allowbreak ranking}, \lean{infinite\_\allowbreak loop\_\allowbreak diverges}, \lean{linear\_\allowbreak recursion\_\allowbreak bigstep}, \lean{binary\_\allowbreak recursion\_\allowbreak bigstep}, \lean{binary\_\allowbreak recursion\_\allowbreak envelope} & \lean{Examples} \\
\bottomrule
\end{tabular}
\addtocounter{table}{-1}\renewcommand{\theHtable}{\arabic{table}.cont}
\caption{Correspondence between the paper and the Lean~4 development (continued).}
\end{table}

\paragraph{Artifact.}
The Lean sources are in the directory \lean{formalization} of release v1.0 of the repository
\begin{center}\url{https://github.com/ldallolio/crosscheck}\end{center}
archived at Zenodo, \href{https://doi.org/10.5281/zenodo.24425070}{doi:10.5281/zenodo.24425070}, and at Software Heritage, \href{https://archive.softwareheritage.org/swh:1:rel:70c77563037e6db2a55b48b3ad4963c86a67773f}{\nolinkurl{swh:1:rel:70c77563037e6db2a55b48b3ad4963c86a67773f}}; they build with \lean{lake build} against Lean \lean{v4.16.0} and the matching mathlib release. The file \lean{formalization/formalization.yaml} gives a machine-readable version of Table~\ref{tab:lean}.

\paragraph{Modelling choices.}
Traces are lists of instructions; the pointed domain $S_\bot$ is \lean{Option S}; partial transitions are functions into \lean{Option S}. The big-step semantics of loops, linear recursion and branching recursion are inductive predicates (\lean{WhileBigStep}, \lean{FramedBigStep}, \lean{TreeBigStep}). Ranking potentials are integer-valued, and the bound $\bar{N}$ is the natural number \lean{rankBound}.

\paragraph{Scope of the mechanization.}
The following points are \emph{not} mechanized, and the corresponding statements rest on the proofs given in the text:
\begin{itemize}
    \item the equation-generating rules of Section~\ref{sec:application}, which describe a method of translation rather than a statement (their justification is Section~\ref{sec:optimization}); and the completeness direction of Farkas' Lemma, which is cited;
    \item in Section~\ref{subsec:cost-logic}, the remarks on weights and architectures and on parallel composition, which are not formalized;
    \item in Section~\ref{subsec:concurrency}, the hypothesis that one unit of cost takes one unit of time, which is a property of the target, not a theorem; Theorem~\ref{thm:response-time} is stated, as in Lean, for the schedules of Definition~\ref{def:interrupt-model}, in which no activation is released before the start, and its classical extension through the busy period, like greedy scheduling of task graphs with dependencies, is discussed with references, not stated as a result;
    \item in Section~\ref{subsec:stack}, that the executions of a machine program are described by activation trees with the depths read from its control-flow graph (a property of the processor model); and in Section~\ref{subsec:validation}, that a C function means its mini-C description (with jumps, Definition~\ref{def:minic-jumps}), including the arithmetic facts assumed of quotients, remainders and masks, the reading of unsigned arithmetic over the integers, that no other code writes the tracked global variables while the function runs, and, for assumed contracts, that they do not touch the ghost variables of Theorem~\ref{thm:validated-jumps};
    \item the lexicographic potential of Proposition~\ref{prop:lex-log} is mechanized as a bound on the length of runs and as a pair of premises of the loop rule (\lean{lex\_\allowbreak potential}), but it is not a rule of the cost logic nor a case of the checker: with a constant $K_2$ it is an affine potential of the existing loop case, and with a symbolic $K_2$ it is not covered (the logarithmic one is a case of the checker, reduced to rule (5) of the logic); the checker of Section~\ref{subsec:checker} covers assignments, unknown values, sequences, conditionals, loops with a linear or conjunctive test, calls of earlier functions (no recursion) and assumed contracts, with bounds affine in the parameters;
    \item the analyzer of the companion paper \cite{crosscheckTool}, and the translation from C or C++ source to the certificates of Section~\ref{subsec:checker}; for a validated certificate (Section~\ref{subsec:validation}) the translation of the source into mini-C instead, and the hypotheses under which a C function means its mini-C description. The checker program is compiled from the Lean definitions of the checker and of validation by Lean's compiler, together with a parser of the certificate format that is not verified.
\end{itemize}

\section{Limitations}
\label{sec:limitations}

\paragraph{Cost is not time.}
The bounds of this paper are bounds on a cost norm. With unit weights they count instructions; with weights $\omega$ they bound execution time only insofar as $\omega(a)$ is a valid worst-case duration of instruction $a$ in every context. On processors with caches, pipelines and branch prediction, instruction durations are context-dependent, and obtaining valid weights is the task of micro-architectural analysis \cite{wilhelm2008worst}. CrossCheck provides the loop-bound and path-cost components of a WCET analysis, not a complete WCET analyzer.

\paragraph{Precision.}
Tropical collapsing ignores infeasible paths. The nested-loop envelope is a product of worst cases without amortization. The branching envelope assumes a full tree of maximal depth, which is exponentially pessimistic for unbalanced recursions; size potentials avoid this only when an affine potential exists. Recursive functions are charged the worst-case cost of one activation for every activation.

\paragraph{Completeness.}
Exact solutions are available only for fragments with closed-form iterates. The relaxed analysis depends on finding a ranking potential; the linear template is incomplete for loops that require lexicographic or non-linear arguments, and it depends on externally supplied invariants.

\paragraph{Language coverage.}
The framework covers structured control flow (sequencing, conditionals, while loops, linear and branching recursion of the given schemes) over a store of scalar variables. Unstructured jumps, pointers and aliasing, dynamic allocation, mutual recursion and concurrency are not covered.

\paragraph{Evaluation.}
This paper contains no experimental evaluation; the companion paper \cite{crosscheckTool} measures an implementation on established benchmarks and compares it with other tools.

\section*{Conclusions}

The relation \textit{order-by-composition} ($\sqsubseteq$) has been defined on the free monoid of execution traces $\Sigma^*$ and on computation trees $\mathcal{T}_\Sigma$ as an intensional partial order along which computational cost is monotone. It justifies replacing iterative loops and recursive control flow with constrained optimization problems: the terminal state corresponds to the first iterate, or to the least computation tree, that satisfies the negation of the continuation condition, and we have proved that this minimizer is exactly the result of the standard big-step semantics.

General recurrence systems are undecidable; relaxing exactness into envelopes based on ranking potentials and the tropical semiring yields sound cost bounds. The bounds compose over whole programs, with procedures, recursion and assumed contracts, in a cost logic that is sound for terminating and non-terminating executions, and a linear instance of the logic has certificates that a small checker validates with integer arithmetic only. The foundations, the envelopes, the cost logic, the checker and the examples are machine-checked in Lean~4.

Future work on the theory includes lexicographic potentials as rules of the cost logic and of the checker, amortization beyond the two forms of Proposition~\ref{prop:amortised-loop} (symbolic entry bounds, non-linear costs), logarithmic loops with symbolic entry bounds or early exits, calls and early exits in the checker, the mechanization of the tree order for arbitrary arity, a verified translation from a fragment of C to the certificates, and the treatment of concurrent execution traces.

\section*{Statement on the Use of AI}

Following the recommendations on the responsible release of AI-generated mathematics \cite{agmai2026responsible}, we document how AI was used in this work. The framework, the examples and the original manuscript are due to the author. In November 2026 an AI coding assistant (Anthropic's Claude, models Opus~5.5 and Fable~5.2, and Haiku~5.5 inside the component that reads web pages, used through the Claude Code tool) was asked to review the manuscript critically and to improve the paper, the bibliography and the Lean development. In that process the assistant restructured and rewrote substantial parts of the text, corrected several statements, compiled the related-work bibliography, and wrote the Lean modules \lean{WhileLoop}, \lean{Ranking}, \lean{Recursion}, \lean{Certificate}, \lean{Arity}, \lean{Language}, \lean{Checker}, \lean{MiniC}, \lean{MiniCJump}, \lean{Interrupts}, \lean{ForkJoin}, \lean{Stack}, \lean{Norms}, \lean{Branching}, and \lean{Divergence} (with the cost logic of Section~\ref{subsec:cost-logic}, the certificates and checker of Section~\ref{subsec:checker}, the validation of Section~\ref{subsec:validation}, the results of Sections~\ref{subsec:concurrency} and~\ref{subsec:stack}, Proposition~\ref{prop:lex-log}, and their statements and proofs in the text) together with extensions of the other modules. The assistant also proposed the simplifications and the bibliography changes of the present version, which the author approved, and wrote the Clang plugin and the evaluation of the companion paper \cite{crosscheckTool}; the prompts and a summary of the changes are recorded in the file \lean{AI\_USAGE.md} of the repository. The assistant also compared the sentences that cite a paper with the copies of the cited works that it could obtain. Every Lean proof is checked by the Lean kernel; the statements that are not mechanized are listed in Section~\ref{sec:mechanization}. An initial draft of the paper was hand-written, including a java program for the mechanical generation and solving of the minimization problem from source code; I have read the text and take full responsibility for the content.

\section*{Acknowledgements}

Thanks are also due to Prof. E. Denti, Dr. B. Franceschiello and Prof.
A. Omicini for their valuable suggestions and discussions.

\bibliographystyle{unsrt}
\bibliography{crosscheck-theory}

\end{document}